\documentclass[aps,prd,superscriptaddress,12pt,showpacs,notitlepage]{revtex4}
	\usepackage{graphicx}
	\usepackage{amsmath,amssymb,mathrsfs}
\usepackage{epsf}
\usepackage{epsfig}
\usepackage{subfigure}
\usepackage{amssymb}
\usepackage{amsmath}
\usepackage{bbm}
\usepackage{bbold}
\usepackage{xcolor}

\newcommand{\be}{\begin{equation}}
\newcommand{\ee}{\end{equation}}
\newcommand{\bea}{\begin{eqnarray}}
\newcommand{\eea}{\end{eqnarray}}

\newcommand{\eqn}{\begin{eqnarray}}
\newcommand{\eqnx}{\end{eqnarray}}

\begin{document}
\title{Compact Q-balls and Q-shells in the $CP^N$ type models}

\author{P. Klimas}
\affiliation{Departamento de F\'isica, Universidade Federal de Santa Catarina, Campus Trindade, CEP 88040-900, Florian\'o—polis-SC, Brazil}
\author{L.R. Livramento}
\affiliation{Departamento de F\'isica, Universidade Federal de Santa Catarina, Campus Trindade, CEP 88040-900, Florian\'o—polis-SC, Brazil}
\affiliation{Instituto de F\'isica de S\~ao Carlos; IFSC, Universidade de S\~ao Paulo, Caixa Postal 369, CEP 13560-970, S\~ao Carlos-SP, Brasil}


\begin{abstract}
We show that the $CP^N$ model with odd number of scalar fields and  V-shaped potential possesses  finite energy compact solutions in the form of Q-balls and Q-shells. The solutions were obtained in 3+1 dimensions. Q-balls appears for $N=1$ and $N=3$ whereas Q-shells are present for higher odd values of $N$. We show that energy of these solutions behaves as $E\sim |Q|^{5/6}$, where $Q$ is the Noether charge.

\end{abstract}

\maketitle 
\section{Introduction}

Field-theoretic models that admit soliton solutions became popular in many branches of physics such as cosmology, particle physics, nuclear physics and condensed matter physics. Solitons are very special stable field configurations whose properties are related with conserved quantities. They are usually studied as solutions of some effective classical non-linear field models which are expected to grasp the most relevant physical properties of underlying quantum theory. For instance, the Skyrme model and the Skyrme-Faddeev (SF) model together with theirs  extensions are intensively studied in the context of description of nuclear matter and strong interactions \cite{sf}. 

Another important group of field-theoretic models is formed by the $CP^N$ models i.e. models on a complex projective space \cite{sigma}.  The $CP^N$ models have a close relation with so-called non-linear sigma models which have applications in diverse areas of physics. For instance, the simplest one in this group, the $CP^1$ model, is related with the model describing Heisenberg ferromagnet \cite{heisenberg} and it is defined by the Lagrangian ${\cal L}_0=\frac{1}{2}\partial_{\mu}\vec\phi\cdot\partial^{\mu}\vec\phi$ , where the triplet of real scalar fields $\vec \phi=(\phi^1,\phi^2,\phi^3)$ satisfy constraint $\vec\phi\cdot\vec\phi=1$.  The relation between  non-linear sigma model and the $CP^1$ model is established by the stereographic projection. An important point about ${\cal L}_0$ is that this Lagrangian appears as a part of Lagrangians of the Skyrme model (defined in terms of $\vec\phi\in S^2$ instead of chiral fields $U\in SU(2)$) and of the SF model.  The general  $CP^N$ model is defined by the Lagrangian ${\cal L}_{CPN}=\lambda^2(D_{\mu}{\cal Z})^{\dagger}D^{\mu}{\cal Z}$, where $\lambda^2$ is a dimensional constant and  $D_{\mu}{\cal Z}\equiv\partial_{\mu}{\cal Z}-({\cal Z}^{\dagger}\cdot\partial_{\mu}{\cal Z}){\cal Z}$. The vector ${\cal Z}$ has the form ${\cal Z}=({\cal Z}_1,\ldots, {\cal Z}_{N+1})$ and satisfy the constraint ${\cal Z}^{\dagger}{\cal Z}=1$. A set of independent complex fields is introduced as ${\cal Z}=(u_1,\ldots,u_N,1)/\sqrt{1+u^{\dagger}\cdot u}$. Some more complex models contain the $CP^N$ Lagrangian as one of terms in a total action. For instance, it takes place for the extended SF model with target space $SU(N+1)/SU(N)\otimes U(1)=CP^N$, see \cite{lafklim}.

An existence of topological solutions of the $CP^N$ models is closely related with the homotopy classes $\pi_k(CP^N)$, where $k$ is a dimension of the base space. According to \cite{adda}, planar $k=2$ models on a coset space $G/H$  posses the homotopy class $\pi_2(CP^N)=\pi_1(H)_G$, where $\pi_1(H)_G$ is a subset  of $\pi_1(H)$ formed by closed paths in $H$ which can be contracted to a point in $G$. It follows, that topological charges of the $CP^N$ model are given by $\pi_1(SU(N)\otimes U(1))_{SU(N+1)}$ and they are equal to number of poles of $u_i$ (including those at infinity).  It has been shown that  the $CP^N$ model and the extended SF model with the $CP^N$ target space posses exact topological vortex solutions \cite{lafklim, vortices} and numerical vortex solutions in the models containing potential \cite{vortpot}. Although such vortices were obtained in  3+1 dimensions their topological charge density and the energy density are functions of merely two spatial coordinates.  Unfortunately, there are no such  solutions for  $k=3$  and $N>1$ because the homotopy class $\pi_3(CP^N)$ is trivial (note that  $\pi_3(CP^1)=Z$). It leads to conclusion that models in three spatial dimensions possessing the $CP^N$ target space, where $N>1$, can have only non-topological solutions. A Derrick's scaling theorem \cite{derrick} provides further restrictions on solutions. It implies a non-existence of static solutions in the $CP^N$ model.  Then, an explicit time dependence can be introduced through the Q-ball ansatz i.e. assumption that phases of all complex fields rotate with equal frequencies $\omega$.  Q-ball configurations are given by scalar fields proportional to the factor $e^{i\omega t}$, see, \cite{lee, rosen, werle}.  

Some field-theoretic models with standard quadratic kinetic terms posses Q-ball solutions, however, existence of such solutions require inclusion of a potential term into the Lagrangian. The form of the potential in vicinity of its minimum determines the  leading behaviour  of the scalar field near the vacuum solution. It has been shown in \cite{al1, al3, al2} that there is a class of Q-ball solutions which approach vacuum solution in quadratic manner, however, it requires potential which have non-vanishing left- and right-hand side derivatives at the minimum. In other words, such potential are sharp at the minimum (V-shaped potentials) \cite{Vshaped}. The models with V-shaped potentials leads to equations of motion containing certain discontinuous terms. A typical field-theoretic model with this property is  the signum-Gordon model \cite{al1, al3, SG}. Solutions of such differential equations  have precise mathematical meaning within formalism of generalized functions. They are so-called {\it weak solutions} of differential equations \cite{weak}. Another very characteristic property of models with V-shaped potentials is existence of {\it compactons} i.e. (solitonic) solutions that differ from a vacuum value on a finite subspace of the base space. In other words, they approach to the vacuum at finite distance and do not have infinitely extended tails - typical for better known solitons. In spite of their unusual properties compactons find many applications: from condense matter physics \cite{ros} to nuclear physics \cite{adam} and cosmology \cite{kunz}. In fact, Q-balls presented in \cite{al1, al3, al2} are examples of compactons. A distinct approach to compact Q-balls based on potentials containing fractional powers is presented in \cite{bazeia}.  

In this paper we shall construct some finite energy compact solutions of the $CP^N$ model with V-shaped potential. We are interested in solutions in three spatial dimensions. Motivation for this work comes from observation that  vortex solutions in models with the $CP^N$ target space defined in 3+1 dimensions have infinite  total energy due to infinite length of the vortices (energy per unit of length is finite). As we are interested in solutions with finite total energy, the compact Q-balls are very good candidates. First,  the Q-ball ansatz allows for time dependent fields. Second, a compactness of solutions guarantee that the total energy is given by the integral over a finite spatial region.  In such a case there is no problem with convergence of the integral at spatial infinity. Construction of such solutions in the $CP^N$ model is an important step into searching for similar solutions in effective models like the extended $CP^N$ SF model.

 The paper is organized as follows. In Section II we introduce the model and its parametrization. Section III is devoted to study of compact Q-balls and Q-shells which differ by number of scalar fields. We compute Noether charges for such solutions and study how the energy depends on these charges.  In section IV we present analytic insight into solutions with small amplitude  (the signum-Gordon limit). We obtain an exact solutions for the limit model and compare them with numerical solutions of the complete non-linear model. In the last section we give some final conclusions.

\section{The model}
We shall study a 3+1 dimensional model with the $CP^N$ target space. The $CP^N$ space is a symmetric space \cite{helgason} and it can be written as a coset space $CP^N=SU(N+1)/SU(N)\otimes U(1)$ with the subgroup $SU(N)\otimes U(1)$ being invariant under the involutive automorphism $(\sigma^2=1)$. The $CP^N$ space has a nice parametrization in terms of the principal variable $X$, see \cite{eihenherr, lafolive}, defined as
\be
X(g):=g\sigma(g)^{-1},\qquad\qquad g\in SU(N+1).\label{pv}
\ee
It satisfy $X(gk)=X(g)$ for $\sigma(k)=k$, where $k\in SU(N)\otimes U(1)$. A parametrization of the $CP^N$ model in terms of the variable $X$ is presented in \cite{lafleit} and of the extended SF model in \cite{lafklim}.
The model we consider here is just the $CP^N$ model extended by a potential (non-derivative) term. As we show it below, such term is crucial to have compactons. The model is given by the Lagrangian 
\be
{\cal L}=-\frac{M^2}{2}{\rm Tr}\,(X^{-1}\partial_{\mu}X)^2-\mu^2V(X),\label{lagr1}
\ee
where $M$ has dimension of mass and the potential $V(X)$ shall be specified in further part of the paper. It has been pointed out in \cite{fkz} that term proportional to $M^2$ is just a Lagrangian of the $CP^{N}$ model. 
Note, that parametrization of the $CP^N$ model in terms of principal variable is as good as that in terms of the complex vector ${\cal Z}$ defined in Introduction. The reason why we employ the principal variable $X$ instead of ${\cal Z}$ is that  the extended $CP^N$ SF model \cite{lafklim} has been defined in terms of this variable.  In such approach, a further inclusion of quartic terms (with intention to search for compact Q-balls in the extended $CP^N$ SF model) would be much easier. 

We assume the $(N+1)$-dimensional defining representation in which the $SU(N+1)$ valued group element $g$ is parametrized by the set of complex fields $u_i$ 
\begin{eqnarray}
g\equiv\frac{1}{\vartheta}\left(\begin{array}{cc}
\Delta&iu\\
iu^{\dagger}&1
\end{array}\right),\qquad\qquad \Delta_{ij}\equiv \vartheta\,\delta_{ij}-\frac{u_iu_j^*}{1+\vartheta},\qquad\qquad\vartheta\equiv \sqrt{1+u^{\dagger} \cdot u},\label{parametryzacja}
\end{eqnarray}
which leads to the following form of the principal variable \eqref{pv} 
\begin{eqnarray}
X(g)=g^2=
\left(\begin{array}{cc}
{\mathbb{1}}_{N\times N} & 0 \nonumber \\
0 & -1 \nonumber 
\end{array}\right)
+
\frac{2}{\vartheta^2}\left(\begin{array}{cc}
-u\otimes u^\dagger & iu \nonumber \\
iu^\dagger & 1  
\end{array}\right).
\end{eqnarray}
The Lagrangian \eqref{lagr1} simplifies to the form
\be
{\cal L}=-M^2\eta^{\mu\nu}\tau_{\nu\mu}-\mu^2V,\label{lagr2}
\ee
where $\eta^{\mu\nu}={\rm diag}(1,-1,-1,-1)$ and
\be
\tau_{\nu\mu}:=-4\frac{\partial_{\mu}u^{\dagger}\cdot\Delta^2\cdot\partial_{\nu}u}{(1+u^{\dagger}\cdot u)^2},\qquad{\rm where}\qquad \Delta^2_{ij}=\vartheta^2\delta_{ij}-u_iu^*_j.
\ee
Variation of the Lagrangian with respect to fields $u_i^*$ leads to set of equations of motion. All terms containing second derivatives can be uncoupled with help of inverse of $\Delta^2_{ij}$ which has the form $\Delta^{-2}_{ij}=\frac{1}{1+u^{\dagger}\cdot u}(\delta_{ij}+u_iu_j^*)$. It gives
\begin{align}
\partial_{\mu}\partial^{\mu}u_i-2\frac{(u^{\dagger}\cdot \partial^{\mu}u)\partial_{\mu}u_i}{1+u^{\dagger}\cdot u}+\frac{\mu^2}{4M^2}(1+u^{\dagger}\cdot u)\sum_{k=1}^N\left[(\delta_{ik}+u_iu^*_k)\frac{\delta V}{\delta u_k^*}\right]=0.\label{EL}
\end{align}

In order to construct compacton solitons we have to carefully chose the potential. We know from previous investigations \cite{Vshaped, top} that for theories with the usual kinetic term (first derivatives squared) one needs a potential which possesses a linear approach to the vacuum. For example, we may use the following potential 
\be
V(X)=\frac{1}{2}\left[{\rm Tr}({ \mathbb{1} }-X)\right]^{\frac{1}{2}}=\left(\frac{u^{\dagger}\cdot u}{1+u^{\dagger}\cdot u}\right)^{\frac{1}{2}},\label{pot1}
\ee
which is the $CP^N$ generalization of the  $CP^1$ (or baby-Skyrme) case \cite{AKSW}. The potential 
vanishes at $u_i=0$ {\it i.e.} $X= \mathbb{1}$.  In absence of the Skyrme term the model discussed in \cite{AKSW} became a $2+1$ dimensional $CP^1$ model  with a potential. The model defined by \eqref{lagr1} and \eqref{pot1} is a 3+1 dimensional  model with a V-shaped potential. As it has been already announced in Introduction, among remarkable properties of such models there is  non-vanishing of the first derivative of the potential at the minimum and  existence of compactons. Such compact solutions consist of appropriately matched non-trivial {\it partial solutions} and a constant vacuum solution. By partial solutions we mean solutions which hold only on some compact support. Matching surfaces correspond with borders of compactons. Unlike for differentiable potentials, the constant vacuum solution does not satisfy  equation with nontrivial potential $V$ but rather equation in the model without potential. The existence of constant solutions can be deduced from the form of the energy density. 

In particular, such solutions are almost straightforward in the field-theoretic models which possess a mechanical realization.
For instance, in the case of the signum-Gordon model with a single real scalar field, the potential has the form $V\propto |\phi|$, so for $V\neq 0$ the equation of motion is of the form $\partial_{\mu}\partial^{\mu}\phi\pm 1=0$. Such a model is physically sound because it can be obtained as a continuous limit of a given mechanical system \cite{Vshaped}. Moreover, it became clear from its mechanical realization that $\phi=0$ is a physical configuration that minimizes the energy (vacuum solution). The vacuum solutions obeys equation $\partial_{\mu}\partial^{\mu}\phi=0$ and it can be formally included replacing the equation of motion by  $\partial_{\mu}\partial^{\mu}\phi+{\rm sgn}(\phi)=0$, where ${\rm sgn}(0):=0$. 

In the model considered in this paper the energy density is given by expression
\begin{align}
{\cal H}&:=\frac{\delta{\cal L}}{\delta(\partial_0u_i)}\partial_0u_i+\frac{\delta{\cal L}}{\delta(\partial_0u^*_i)}\partial_0u^*_i-{\cal L}\nonumber\\
&=-M^2\left(\tau_{00}+\sum_{a=1}^3\tau_{aa}\right)+\mu^2V,\label{hamdim}
\end{align}
where the index $a$ labels spatial Cartesian coordinates $x^a$. It vanishes for a constant field configurations $u_i=0$, where $i=1,\ldots,N$. The vacuum configuration satisfies a homogeneous $CP^N$ equation
\begin{align}
\partial_{\mu}\partial^{\mu}u_i-2\frac{(u^{\dagger}\cdot \partial^{\mu}u)\partial_{\mu}u_i}{1+u^{\dagger}\cdot u}=0,\label{eq1}
\end{align}
whereas a non-constant partial solution must satisfy the equation following from \eqref{EL}
\begin{align}
\partial_{\mu}\partial^{\mu}u_i-2\frac{(u^{\dagger}\cdot \partial^{\mu}u)\partial_{\mu}u_i}{1+u^{\dagger}\cdot u}+\frac{\mu^2}{8M^2}\frac{u_i}{\sqrt{u^{\dagger}\cdot u}}\sqrt{1+u^{\dagger}\cdot u}=0.\label{eq2}
\end{align}

The parametrization \eqref{parametryzacja} fixes the global  $U(N+1)$ symmetry of the model to $SU(N)\otimes U(1)$. Its subgroup $U(1)^N$ is given by a set of transformations 
\be
u_i\rightarrow e^{i\alpha_i}u_i,\qquad\qquad i=1,2,\ldots,N,\label{symmtr}
\ee
where $\alpha_i$ are some global continuous parameters.
Symmetry transformation \eqref{symmtr} of the model leads to conserved Noether currents  
\be
J^{(i)}_{\mu}=-\frac{4i M^2}{(1+u^{\dagger}\cdot u)^2}\sum_{j=1}^N\left[u^*_i\Delta^2_{ij} \partial_{\mu}u_j-\partial_{\mu}u^*_j\Delta^2_{ji}u_i\right].\label{prady1}
\ee
The Noether currents \eqref{prady1} satisfy the continuity equation $\partial^{\mu}J^{(i)}_{\mu}=0$. If spatial components of  currents \eqref{prady1} vanish at spatial infinity then integration of this equation on the region of spacetime $[t',t'']\times {\mathbb R^3}$ leads to conserved charges 
\be
Q_0^{(i)}=\int_{{\mathbb R^3}} d^3 J^{(i)}_{0}.\label{Q0}
\ee
The charges \eqref{Q0} are fundamental quantities in analysis of stability of non-topological solutions. They constitute the set of additive conserved quantities. If there is known relation between the energy of solutions and the Noether charges  then one can evaluate whether splitting of solution into smaller pieces is energetically favorable or not.

\section{Nontopological solutions of the $CP^{2l+1}$ type model}

We shall restrict our consideration to the case of odd $N$. The spherical harmonics form the finite representation of eigenfunctions of angular part of Laplace's operator. In such a case each complex field can be chosen as proportional to one of $N=2l+1$ spherical harmonics labeled by $l=0,1,\ldots$.  In present paper we shall deal with the $CP^{2l+1}$ target space. For further convenience we shall label $2l+1$ complex fields by $u_{-l},\ldots, u_l$ instead of  $u_1, u_2, \ldots u_{2l+1}$.

It is convenient to parametrize the model in terms of dimensionless coordinates. They can be defined in following way $\tilde  x^{\mu}:=r_0^{-1}x^{\mu}$, where $r_0$ is a constant parameter with dimension of length. Such a constant can be chosen as inverse of dimensional coupling constant $M$ {\it i.e.}  $r_0\equiv M^{-1}$. We consider the ansatz
\be
u_{m}(t,r,\theta,\phi)=\sqrt{\frac{4\pi}{2l+1}}f(r)Y_{lm}(\theta,\phi)e^{i\omega t},\label{ansatz}
\ee
where all coordinates $(t, r,\theta,\phi)$ are dimensionless. They are defined in the following way 
\begin{align}
\tilde x^0=t,\qquad \tilde x^1=r\sin\theta\cos\phi,\qquad \tilde x^2=r\sin\theta\sin\phi, \qquad \tilde x^3=r\cos\theta.
\end{align}
The integer number $l$ is fixed for a given $CP^{2l+1}$ model whereas the index $m$ takes values $-l\le m\le l$.  Taking into account that $\sum_{m=-l}^lY^*_{lm}(\theta,\phi)Y_{lm}(\theta,\phi)=\frac{2l+1}{4\pi}$ we include the factor  $\sqrt{\frac{4\pi}{2l+1}}$ in \eqref{ansatz} to simplify formulas. It follows that $u^{\dagger}\cdot u=f^2(r)$ depends only on the radial coordinate $r$. Similarly, many other  terms either vanish or depend only on the radial coordinate, see Appendix. The field equations \eqref{eq1} and \eqref{eq2} result in a single ordinary differential equation 
\begin{align}
f''+\frac{2}{r}f'+\omega^2\frac{1-f^2}{1+f^2}f-\frac{l(l+1)}{r^2}f-\frac{2ff'^2}{1+f^2}=\frac{\tilde \mu^2}{8}{\rm sgn}(f)\sqrt{1+f^2},\label{radialeq}
\end{align}   
where $\tilde\mu^2:=\mu^2/M^4$ and where we have adopted definition of the ${\rm sgn}()$ function such that ${\rm sgn}(f):=1$ for $f>0$ and ${\rm sgn}(0):=0$. This is the main equation we will further analyze in the paper.

The hamiltonian density \eqref{hamdim} can be written as ${\cal H}=M^4H$, where $H$ is a dimensionless expression
\begin{align}
H=-\left[\tau_{tt}+\tau_{rr}+\frac{1}{r^2}\left(\tau_{\theta\theta}+\frac{1}{\sin^2\theta}\tau_{\phi\phi}\right)\right]+\tilde\mu^2V.\nonumber
\end{align}
For the class of solutions given by \eqref{ansatz} we get a dimensionless energy density which is a function of the radial coordinate itself
\begin{align}
H=\frac{4}{(1+f^2)^2}\left[f'^2+\left(\omega^2+\frac{l(l+1)}{r^2}(1+f^2)\right)f^2\right]+\tilde\mu^2\frac{f\,{\rm sgn}(f)}{\sqrt{1+f^2}}.\label{hamdens}
\end{align}
A total  dimensionless energy is given by the integral
\be
E=\int_{\mathbb{R}^3} d\Omega\, dr\,r^2\,H=4\pi\int_{0}^{\infty}\, dr  r^2H.\label{energia1}
\ee


Let us consider the Noether currents \eqref{prady1}. Since in dimensionless Cartesian coordinates $\tilde x^{\mu}$ the partial derivatives are of the form $\frac{\partial}{\partial x^{\mu}}=M\frac{\partial}{\partial \tilde x^{\mu}}$ then one can define expressions $\tilde J^{(m)}_{\mu}$ as dimensionless quantities  $ J^{(m)}_{\mu}=M^3\tilde J^{(m)}_{\mu}$, where the index $i=1,\ldots,2l+1$  has been replaced by the index  $m=-l,\ldots,l$. The complex fields $u_m$ are functions of curvilinear dimensionless coordinates $\xi^{\mu}\rightarrow\{t,r,\theta,\phi\}$. The Noether currents expressed in these coordinates {\it i.e.} $\tilde J^{(m)}_{\mu}(\xi)$ must satisfy the continuity equation 
\be
\frac{1}{\sqrt{-g}}\partial_{\mu}\left(\sqrt{-g}g^{\mu\nu}\tilde J^{(m)}_{\nu}(\xi)\right)=0,\nonumber
\ee
where $g^{\mu\nu}={\rm diag}(1,-1,-\frac{1}{r^2},-\frac{1}{r^2\sin^2\theta})$ and $\sqrt{-g}=r^2\sin\theta$.  It turns out that there are only two non-vanishing components of the Noether currents, namely
\begin{align}
\tilde J^{(m)}_t(r,\theta)&=8\,\omega\,\frac{(l-m)!}{(l+m)!}\frac{f^2}{(1+f^2)^2}(P_l^m(\cos\theta))^2,\label{currt}\\
\tilde J^{(m)}_\phi(r,\theta)&=8\,m\,\frac{(l-m)!}{(l+m)!}\frac{f^2}{1+f^2}(P_l^m(\cos\theta))^2.\label{currphi}
\end{align}
Note, that both non-vanishing components do not depend on $t$ neither $\phi$.  It follows that   the continuity equation $\partial_tJ^{(m)}_{t}+\frac{1}{r^2\sin^2\theta}\partial_{\phi}J^{(m)}_{\phi}=0$ is satisfied explicitly. The Noether charges can be obtained integrating the continuity equation in the region $[t',t'']\times {\mathbb R^3}$
\be
\int_{t'}^{t''}dt\int_{\mathbb R^3}d^3\xi\left[\sqrt{-g}\partial_{t}\tilde J^{(m)}_t+\partial_a(\sqrt{-g}g^{ab}\tilde J^{(m)}_b)\right]=0,
\ee
where $a,b=\{1,2,3\}$. The second term can be written as a surface integral at spatial infinity and it gives no contribution to the integral if $\tilde J^{(m)}_b$ vanish sufficiently quickly at spatial infinity. The remaining term express equality of the Noether charges
\begin{align}
Q^{(m)}_t&:=\frac{1}{2}\int_{\mathbb R^3} \,d^3\xi\sqrt{-g}\,\tilde J^{(m)}_t(\xi)\label{Noethercharge}
\end{align}
at $t'$ and $t''$. The factor $\frac{1}{2}$ has been introduced for further convenience. Plugging  \eqref{currt} into \eqref{Noethercharge} we get
\begin{align}
Q^{(m)}_t=\omega\,\frac{16\pi}{2l+1}\int_0^{\infty}dr r^2\frac{f^2}{(1+f^2)^2}.\label{Noethercharge2}
\end{align}
All the Noether charges have the same value because \eqref{Noethercharge2} does not depend on $m$. Notice, that contribution to total energy which has origin in $\tau_{tt}$ (proportional to $\omega^2$) can be expressed in terms of the Noether charges as a sum $\sum_{m=-l}^l\omega\, Q^{(m)}_t$. In fact, contributions to the energy which have origin in terms $\tau_{\theta\theta}$ and $\tau_{\phi\phi}$ can also be represented in form of the sum. We have already seen that spatial components of the Noether currents do no contribute to the charges, however, they are useful to define the integrals
\begin{align}
Q^{(m)}_\phi&:=\frac{3}{2}\int_{\mathbb R^3} d^3 \xi\sqrt{-g}\,\,\frac{\tilde J^{(m)}_\phi(\xi)}{r^2}.\label{integrals}
\end{align}
Plugging \eqref{currphi} into \eqref{integrals} and making use of standard orthogonality relation for the associate Legendre functions $\int_{-1}^1dx (P_l^m(x))^2=\frac{2}{2l+1}\frac{(l+m)!}{(l-m)!}$ we get
\be
Q^{(m)}_\phi=m\,\frac{48\pi}{2l+1}\int_0^{\infty}dr \frac{f^2}{1+f^2}.
\ee
The total energy $E$  can be expressed in terms of the Noether charges $Q^{(m)}_t$ and the integrals $Q^{(m)}_\phi$ and it reads
\be
E=4\pi\int_0^{\infty} dr r^2\left(\frac{4f'^2}{(1+f^2)^2}+\tilde\mu^2\frac{f\,{\rm sgn}(f)}{\sqrt{1+f^2}}\right)+\sum_{m=-l}^l\left(\omega\, Q^{(m)}_t+m\,Q^{(m)}_\phi\right),
\ee
where we made use of expression $\sum_{m=-l}^lm^2=\frac{1}{3}l(l+1)(2l+1)$.
In the subsequent part of the paper we shall construct non-topological compact solutions of eq. \eqref{radialeq}.

\subsection{Expansion at the center}
Plugging the series expansion of $f(r)$ at $r=0$ 
\be
f(r)=\sum_{k=0}^{\infty}a_kr^k\label{expandsol}
\ee
to the equation \eqref{radialeq} we get 
\be
\sum_{k=0}^{\infty}b_kr^{k-2}=0,\label{expandeq}
\ee
where the lowest  three coefficients $b_0$, $b_1$, $b_2$ have the form
\begin{align}
b_0&=l(l+1)a_0,\nonumber\\
b_1&=(l-1)(l+2)a_1,\nonumber\\
b_2&=(l-2)(l+3)a_2+\frac{\mu^2}{8}\sqrt{1+a_0^2}+
\left[\frac{2(a_1^2+a_0^2\omega^2)}{1+a_0^2}-\omega^2\right]a_0.\nonumber
\end{align}
The equation \eqref{expandeq} is fulfilled if $a_k$ are such that all coefficients $b_k$ vanish. It turns out that the  form of expansion \eqref{expandsol} is sensitive on the value of the number $l$ {\it i.e.} the number of complex scalar fields $u_m$, where $m=-l\ldots l$. In the following part we study some qualitatively different forms of expansion for $l=0$, $l=1$ and $l\ge2$.

\subsubsection{Case $l=0$}

For $l=0$ the equation  \eqref{expandeq} can be satisfied in the leading term of expansion if $a_1=0$. A more detailed study shows that the choice $a_1=0$ implies vanishing of all odd-order coefficients $a_{2j+1}$, where $j=1,2,3,\ldots$. Indeed, one can check that if for some fixed odd number $n$ all odd lower-order  coefficients vanish $a_1=a_3=\ldots =a_{n-2}=0$ (and consequently vanish all odd-order coefficients up to $b_{n-2}$), then the next odd-order coefficient $b_n$ is of the form $b_n=a_n(l-n)(l+1+n)$. Consequently, in order to put $b_n=0$, one has to set $a_n=0$.  All even coefficients $a_{2j}$ are uniquely determined by $a_0$ which is a free parameter of expansion. In the lowest order of expansion the function $f(r)$, given by \eqref{expandsol},  reads
\be
f(r)=a_0+\left[\frac{\tilde\mu^2}{48}\sqrt{1+a_0^2}-\frac{a_0(1-a_0^2\omega^2)}{6(1+a_0^2)}\right]r^2+{\cal O}(r^4).
\ee
A value of the coefficient $a_0$ can be determined only for a complete solution that must be regular at the center and  at the boundary.
\begin{figure}[h!]
\centering
\subfigure{\includegraphics[width=0.35\textwidth,height=0.25\textwidth, angle =0]{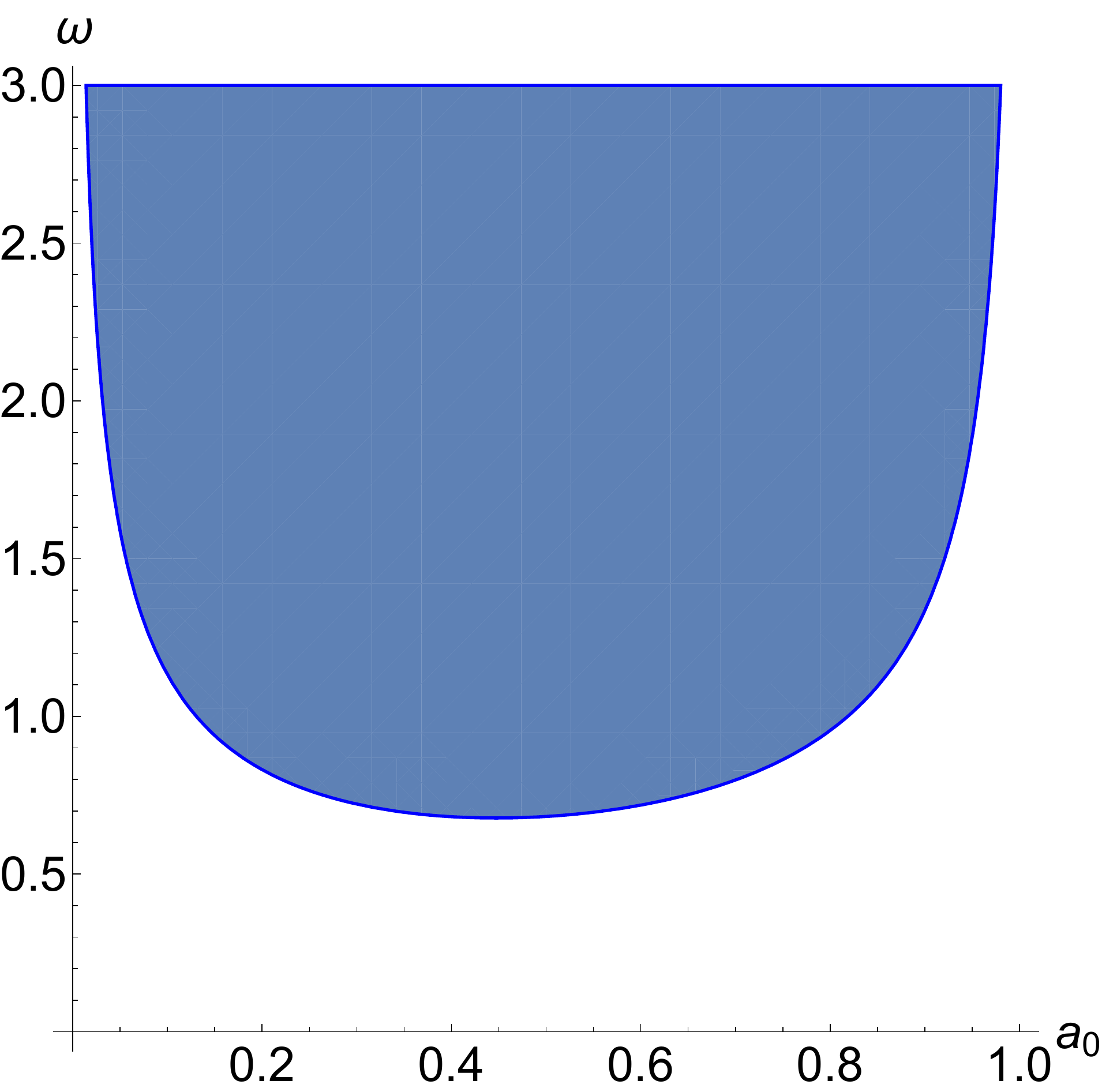}}
\caption{The region $a_2<0$ in dependence on $a_0$ and $\omega$ for the case of $l=0$ where in addition we set $\tilde\mu=1$.}\label{bound}
\end{figure}
A numerical analysis show that physically relevant solution must have $a_2<0$, otherwise the solution grows up infinitely with $r$.  The coefficient $a_2$ depends on the parameter $\omega$. It turns out, that there is a lower bound for $\omega$. The region $a_2<0$ has been sketched in  Fig.\ref{bound}. It suggest the existence of a minimal value of $\omega_{m}$ for which there still exist a compact solution with finite energy. Note, that the border of the region plotted in Fig.\ref{bound} does not determine a value $\omega_m$ but it rather constitutes its limitation. The value of $\omega_m$ can be obtained performing numerical integration of the radial equation \eqref{radialeq}.
It follows from expansion of \eqref{hamdens} at $r=0$
\be
H(r)=a_0\left[\frac{\tilde\mu^2}{\sqrt{1+a_0^2}}+\frac{4a_0\omega^2}{(1+a_0^2)^2}\right]+{\cal O}(r^2),
\ee
that the energy density for $l=0$ does not vanish at the center of Q-ball.

\subsubsection{Case $l=1$}
For $l=1$ the coefficient $a_1$ became a free parameter whereas $a_0$ must vanish. The coefficient $a_2$ is determined by the strength coupling constant $\tilde\mu^2$. Except $a_2$, all higher-order coefficients $a_k$ contain $a_1$. Next three  coefficients of expansion read
\be
a_2=\frac{\tilde\mu^2}{32},\qquad a_3=\frac{a_1}{10}(2a_1-\omega^2),\qquad a_4=\frac{\tilde\mu^2}{576}(12a_1^2-\omega^2).
\ee
Although the radial function satisfies $f(r=0)=0$, the energy density is still non-zero at the center. It can be seen from 
\be
H(r)=12a_1^2+2\tilde\mu^2a_1\,r+\left[\frac{7}{128}\tilde\mu^4-8a_1^4\right]r^2+{\cal O}(r^3).
\ee

\subsubsection{Case $l\ge2$}
It follows from the expansion \eqref{expandeq} that for $l=2,3,\ldots$ both coefficients $a_0$ and $a_1$ must vanish. Taking $a_0=a_1=0$ we do not get non-trivial solution because vanishing of $a_0$ and $a_1$ leads to $a_k=0$ for $k=2,3,\ldots$. It follows that there is no solution which is non-vanishing in the vicinity of $r=0$. However, it does not mean that there are no solutions at all. The radial function cannot be non-trivial at the center but it can be nontrivial at some region  $r\in (R_1,R_2)$. Outside this region {\it i.e.} at $r\in[0,R_1]$ and at $r\in[R_2,\infty)$ the function $f(r)$ vanishes identically. It means that the solution has the form of compact spherical shell. Discussion of behaviour of the radial function $f(r)$ at inner $R_1$ and outer $R_2$ radius is essentially the same.  It is a subject of the next paragraph.


\subsection{Expansion at the boundary}
As we consider compact solutions, the vacuum solution $f(r)=0$ holds for $r> R$, what leads to vanishing of the energy density in this region. A symbol $R$ stands for the compacton radius in the case $l=0,1$ and the outer compacton radius $R\equiv R_2$ for $l=2,3,\ldots$. The continuity of the energy density imposes conditions on the leading behaviour of the solution in the region $r\le R$. Such solution must satisfy following conditions at the border
\be
f(R)=0,\qquad\qquad f'(R)=0.
\ee
Plugging expression $f(r)=A (R-r)^{\alpha}+\ldots$ into \eqref{radialeq} one can find 
\be
-A\alpha(\alpha-1)(R-r)^{\alpha-2}+\frac{\tilde\mu^2}{8}+\ldots=0.\label{leading1}
\ee
The leading term of \eqref{leading1} vanishes for $\alpha=2$ and appropriate value of $A$. It suggest that solutions posses quadratic leading behaviour at the border 
\be
f(r)=\sum_{k=2}^{\infty}A_k(R-r)^k.
\ee
It turns out, that all coefficients $A_k$ are determined in terms of the compacton radius $R$ and parameters of the model. The lowest three coefficients read
\be
A_2=\frac{\tilde\mu^2}{16},\qquad A_3=\frac{\tilde\mu^2}{24\,R},\qquad A_4=\frac{\tilde\mu^2}{192\,R^2}[l(l+1)+8-R^2\omega^2].
\ee
It leads to the following expansion of the energy density at the compacton boundary 
\be
H(r)=\frac{\tilde \mu^2}{8}(R-r)^2+\frac{\tilde \mu^4}{6\,R}(R-r)^3+\frac{\tilde \mu^4}{96\,R^2}\left[4l(l+1)+26-R^2\omega^2\right](R-r)^4+\ldots.
\ee 
The lowest order terms do not depend on the integer number $l$ {\it i.e.} they have the same form independently on the number of complex scalar fields. A first term which depends on $l$ is proportional to $(R-r)^4$.

For  the case $l\ge 2$, {\it i.e.} when the solution has the form of shell-shape compacton, the radial function possesses expansion $f(r)=B_2(r-R_1)^2+\ldots$ at the inner compacton radius $R_1$. The expansion coefficients are almost the same as for the outer compacton radius and they read $B_k=(-1)^kA_k$.

\subsection{Numerical solutions}
\begin{figure}[h!]
\centering
\subfigure[]{\includegraphics[width=0.3\textwidth,height=0.15\textwidth, angle =0]{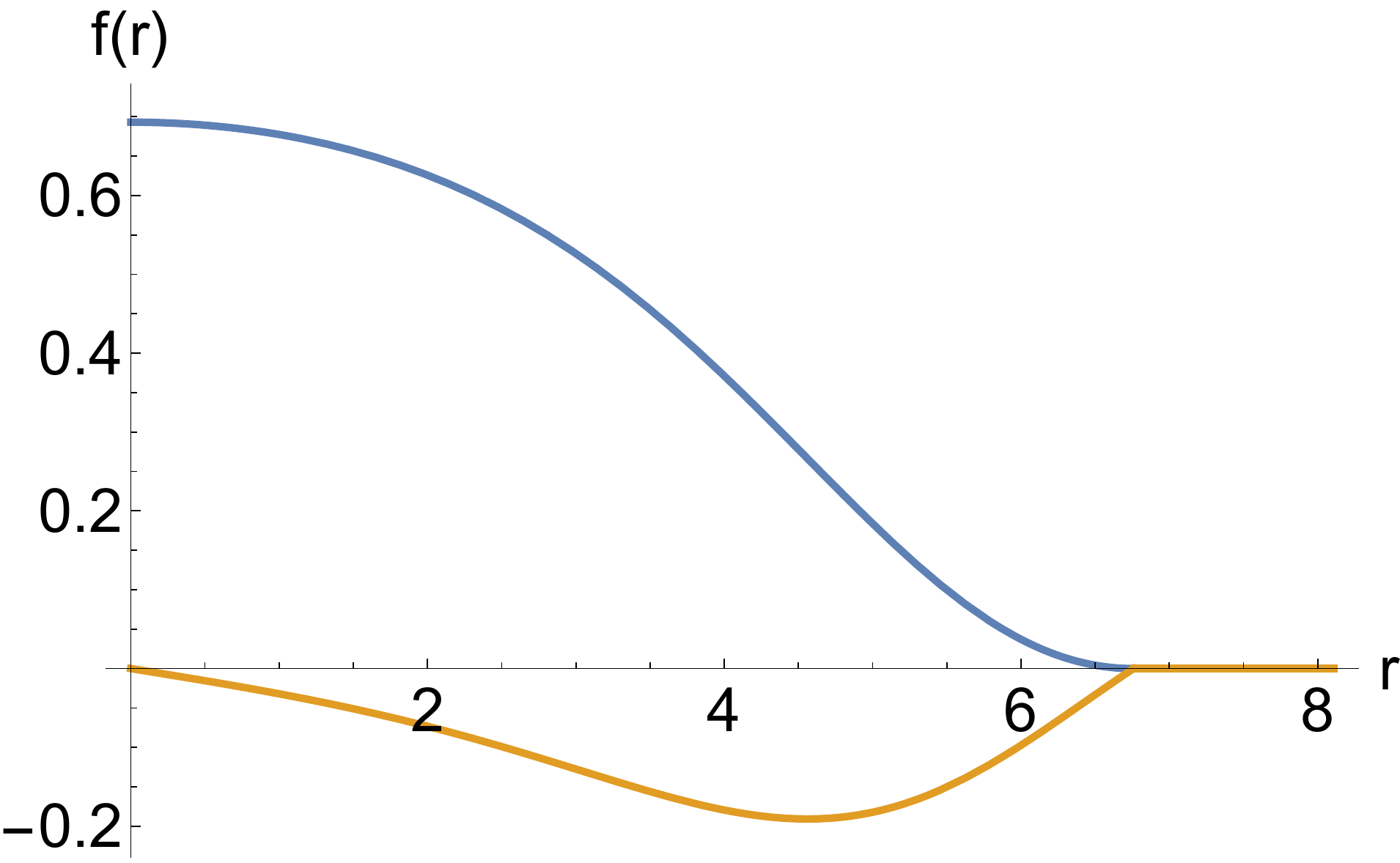}}
\subfigure[]{\includegraphics[width=0.3\textwidth,height=0.15\textwidth, angle =0]{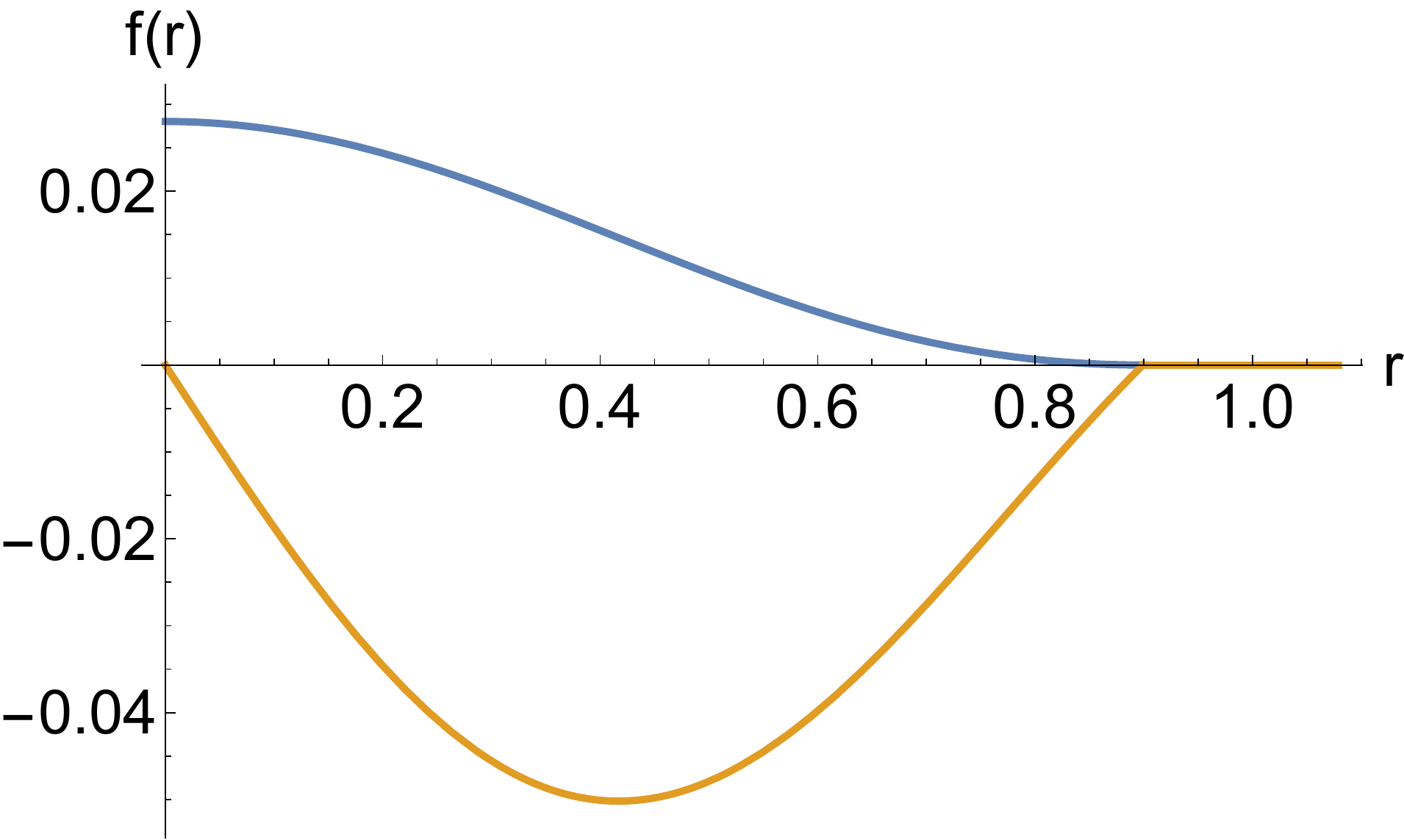}}
\subfigure[]{\includegraphics[width=0.3\textwidth,height=0.15\textwidth, angle =0]{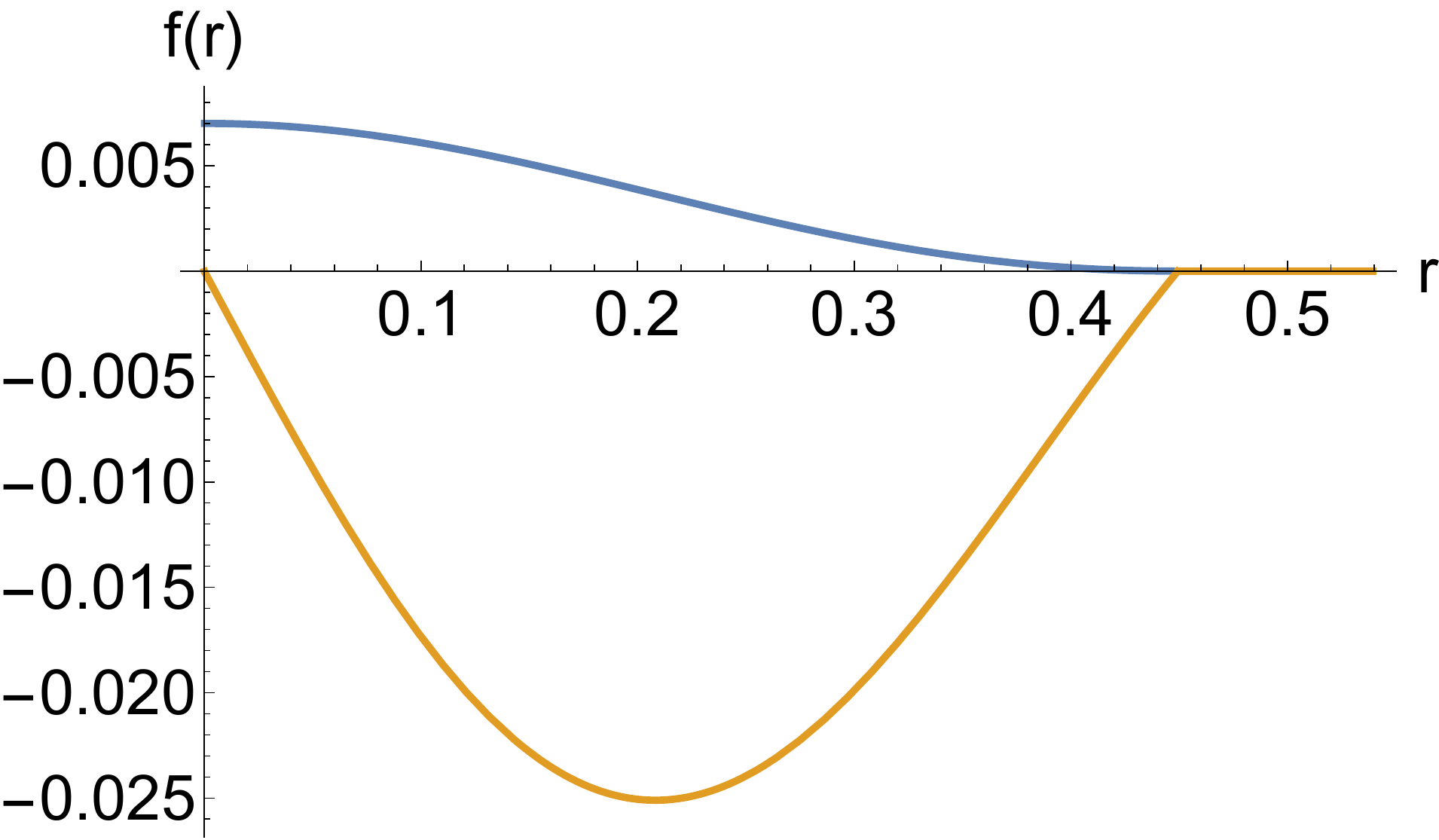}}
\subfigure[]{\includegraphics[width=0.3\textwidth,height=0.15\textwidth, angle =0]{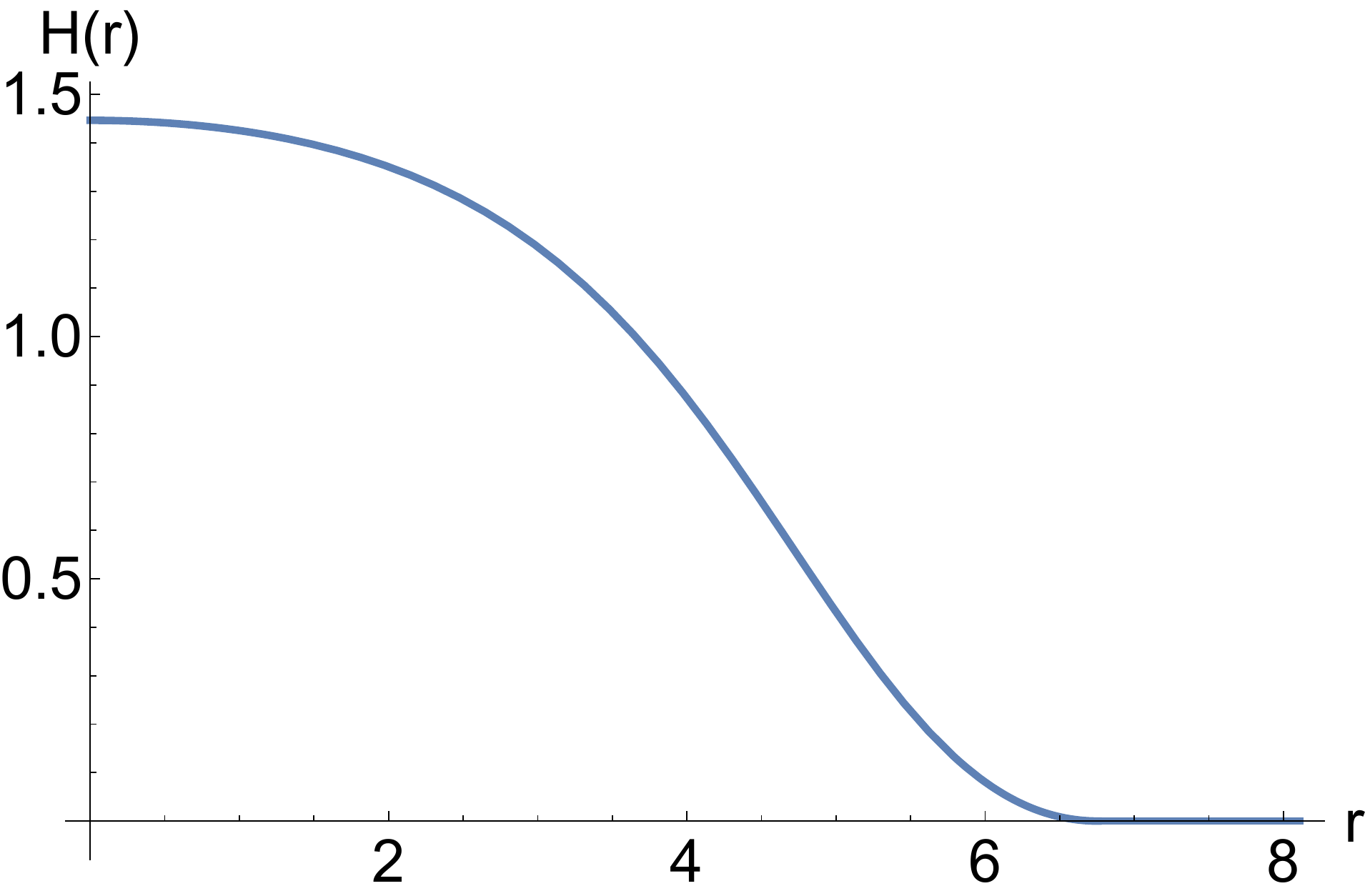}}
\subfigure[]{\includegraphics[width=0.3\textwidth,height=0.15\textwidth, angle =0]{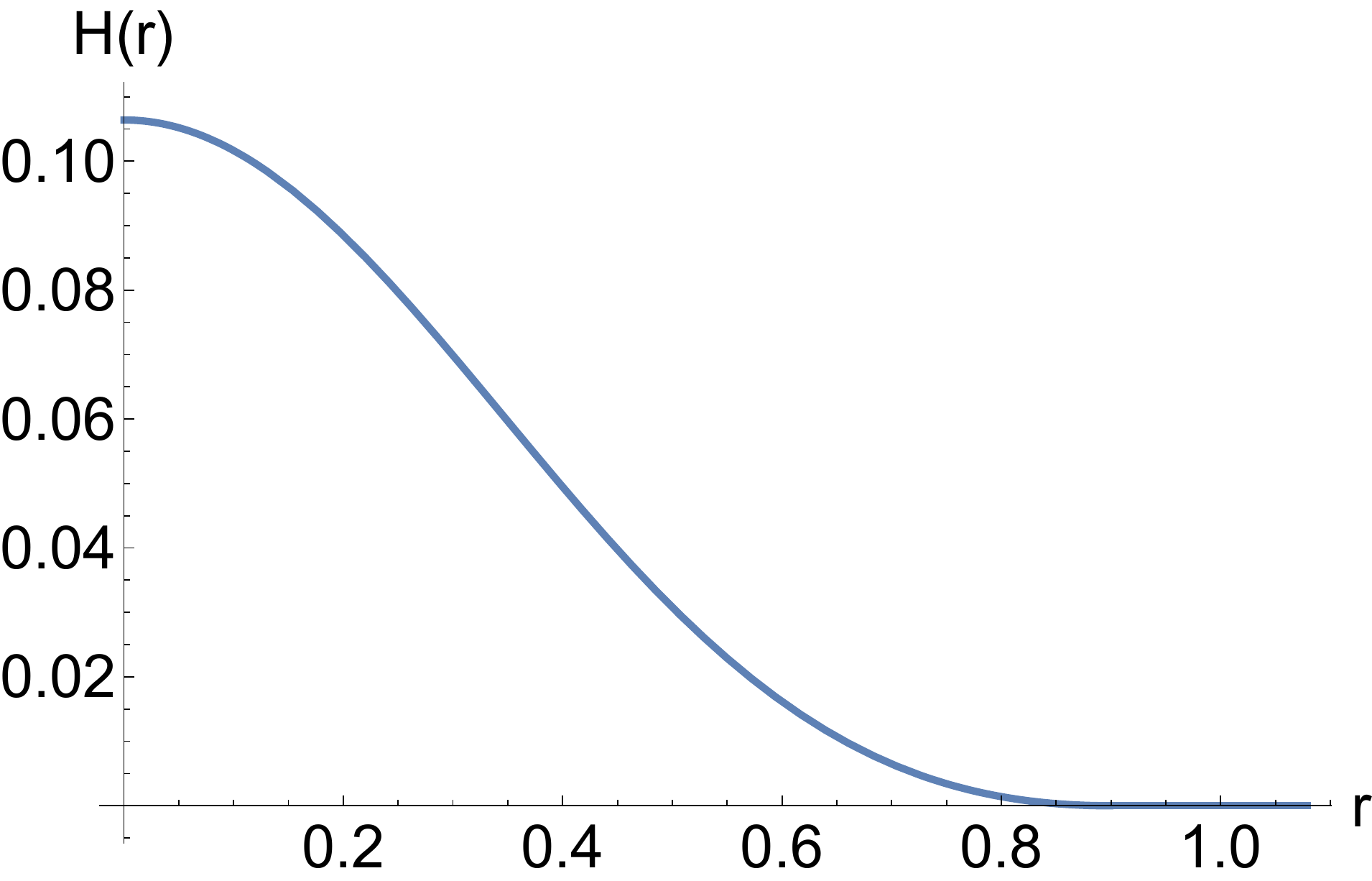}}
\subfigure[]{\includegraphics[width=0.3\textwidth,height=0.15\textwidth, angle =0]{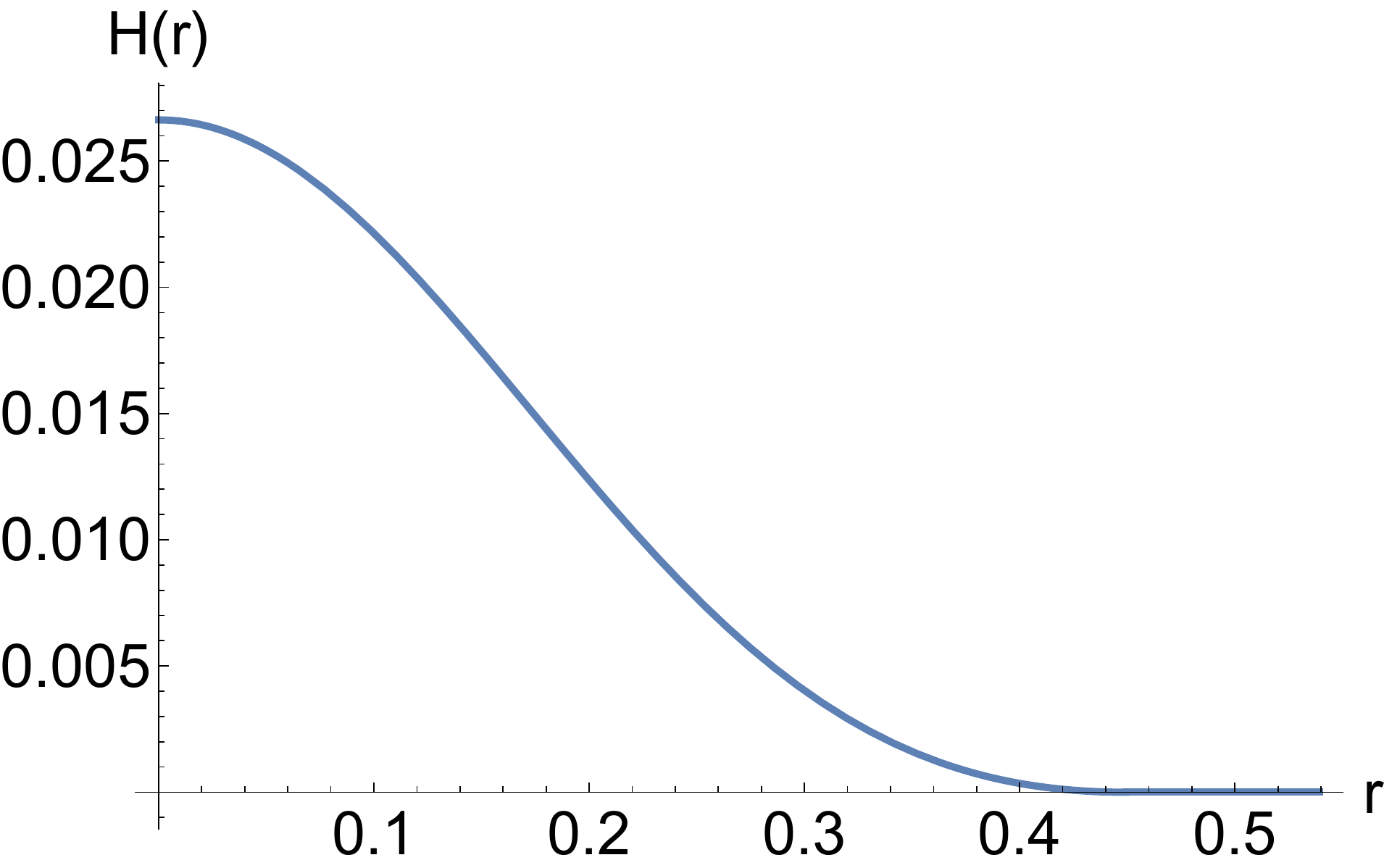}}
\caption{The radial function $f(r)$, its derivative $f'(r)$ and the energy density $H(r)$ for $l=0$ and (a,d) $\omega=1.0$, (b,e), $\omega=5.0$, (c,f) $\omega=10.0$.}\label{l=0}
\end{figure}

We adopt a shooting method for numerical integration of radial equation \eqref{radialeq}. We impose the initial conditions for numerical integration in form of first few terms of series expansion at $r=0$ for $l=0,1$ or $r=R$ for $l\ge2$. In numerical computation we substitute $r=0$  by $r=\varepsilon=10^{-4}$. There is only one free parameter which determine expansion series at the center, namely $a_0$ for $l=0$ and $a_1$ for $l=1$. On the other hand a series expansion at the boundary has also one free parameter, which is the compacton radius $R$. There is exactly one curve being a solution of a second order ordinary differential equation which simultaneously satisfies conditions at the center and at the boundary. For a chosen value of $a_0$ or $a_1$ we integrate numerically the radial equation and determinate a value of the radius $\bar R$ such that $f'(\bar R)=0$. A value of the expression $f(\bar R)$ is used to modify an initial shooting parameter  according to $f(\bar R)\rightarrow 0$ for $\bar R\rightarrow R$. The loop is interrupted when $|f(\bar R)|<10^{-6}$. The examples of numerical solutions for $l=0$ and different values of parameter $\omega$ are presented in Fig.\ref{l=0}.  The compacton profile functions $f(r)$ and their first derivatives $f'(r)$ are sketched in pictures (a), (b) and (c). The respective energy densities are presented in Fig.\ref{l=0} (d), (e), (f). The energy density has maximum at the center $r=0$. 

\begin{figure}[h!]
\centering
\subfigure[]{\includegraphics[width=0.3\textwidth,height=0.15\textwidth, angle =0]{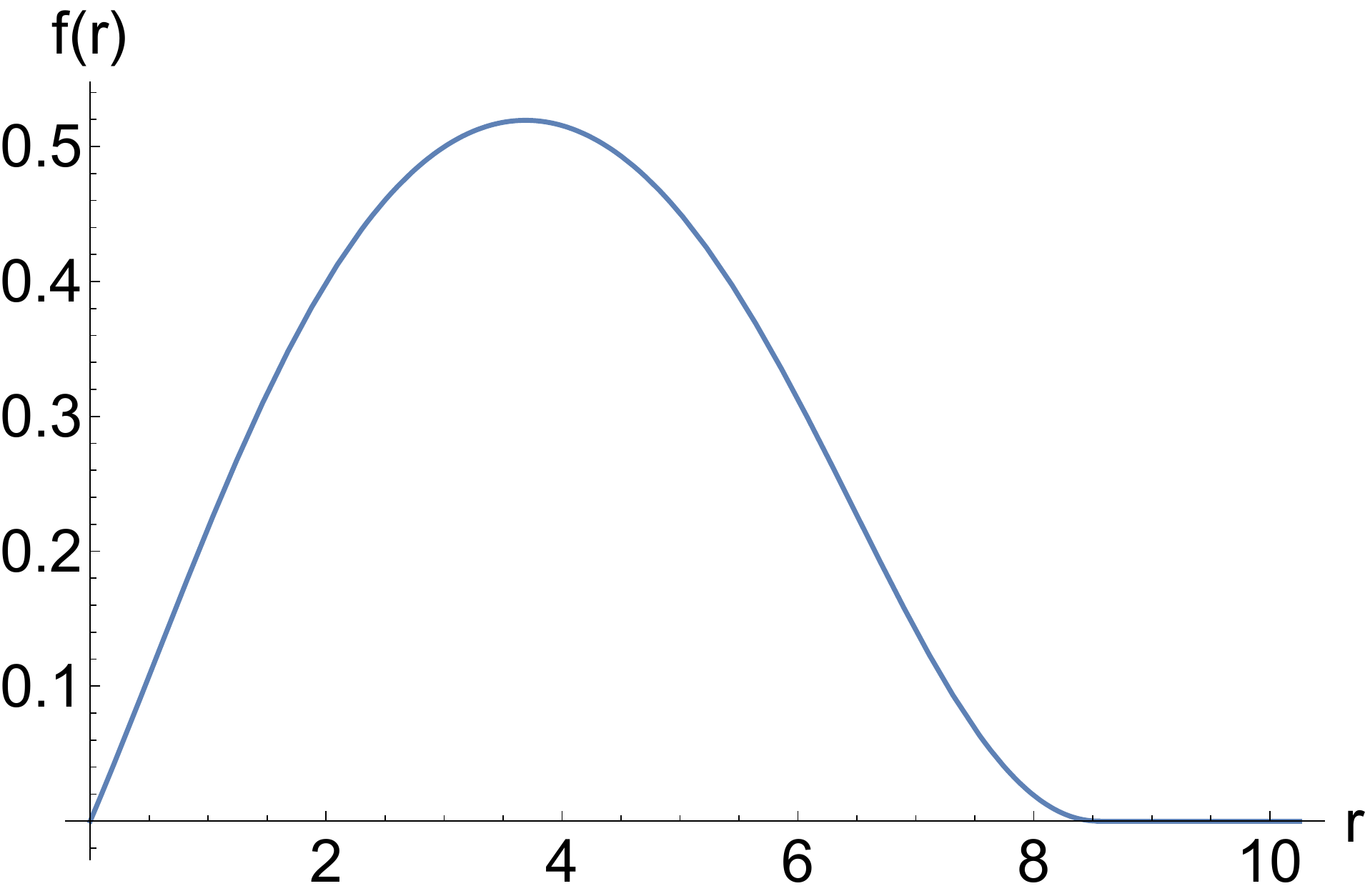}}
\subfigure[]{\includegraphics[width=0.3\textwidth,height=0.15\textwidth, angle =0]{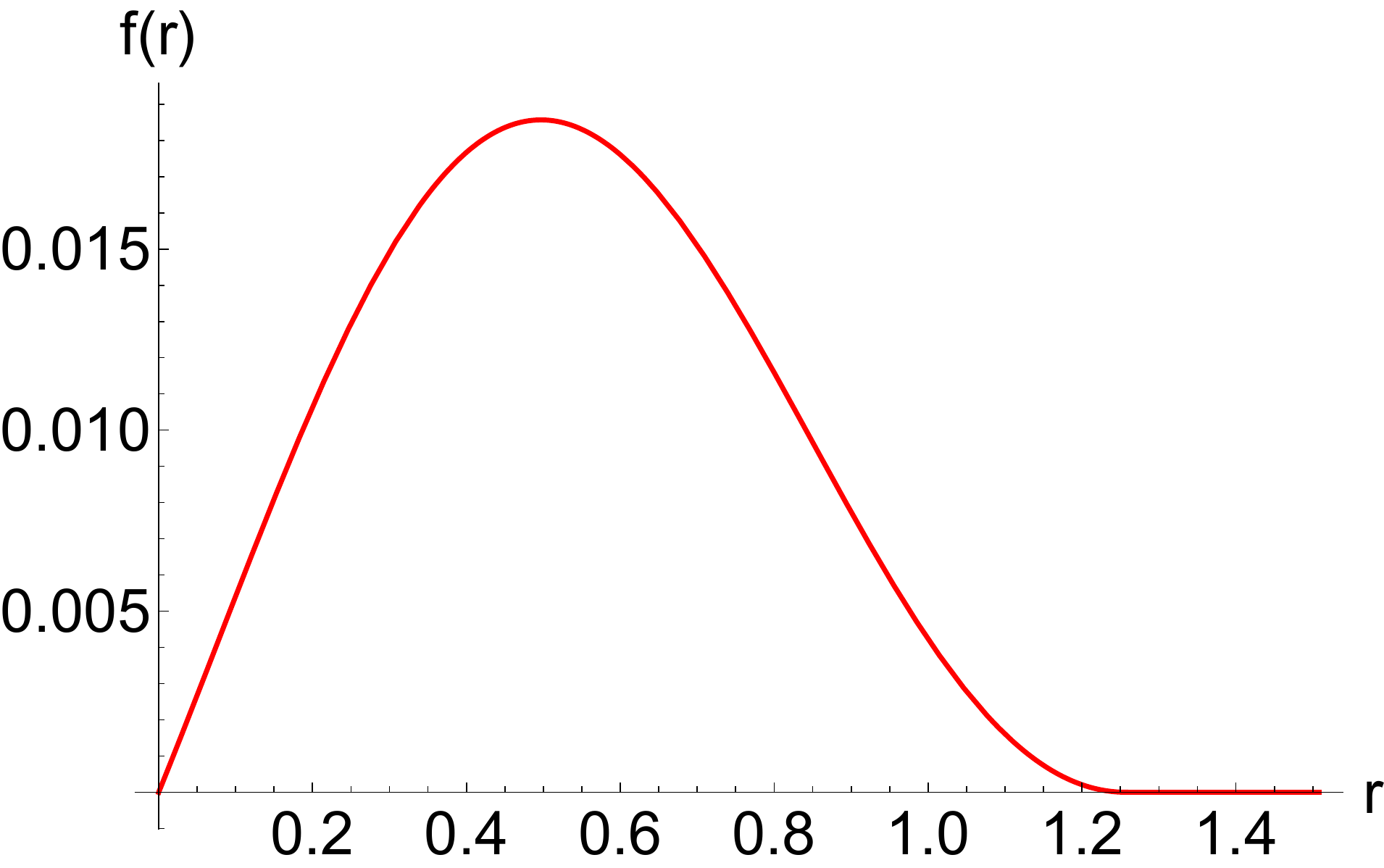}}
\subfigure[]{\includegraphics[width=0.3\textwidth,height=0.15\textwidth, angle =0]{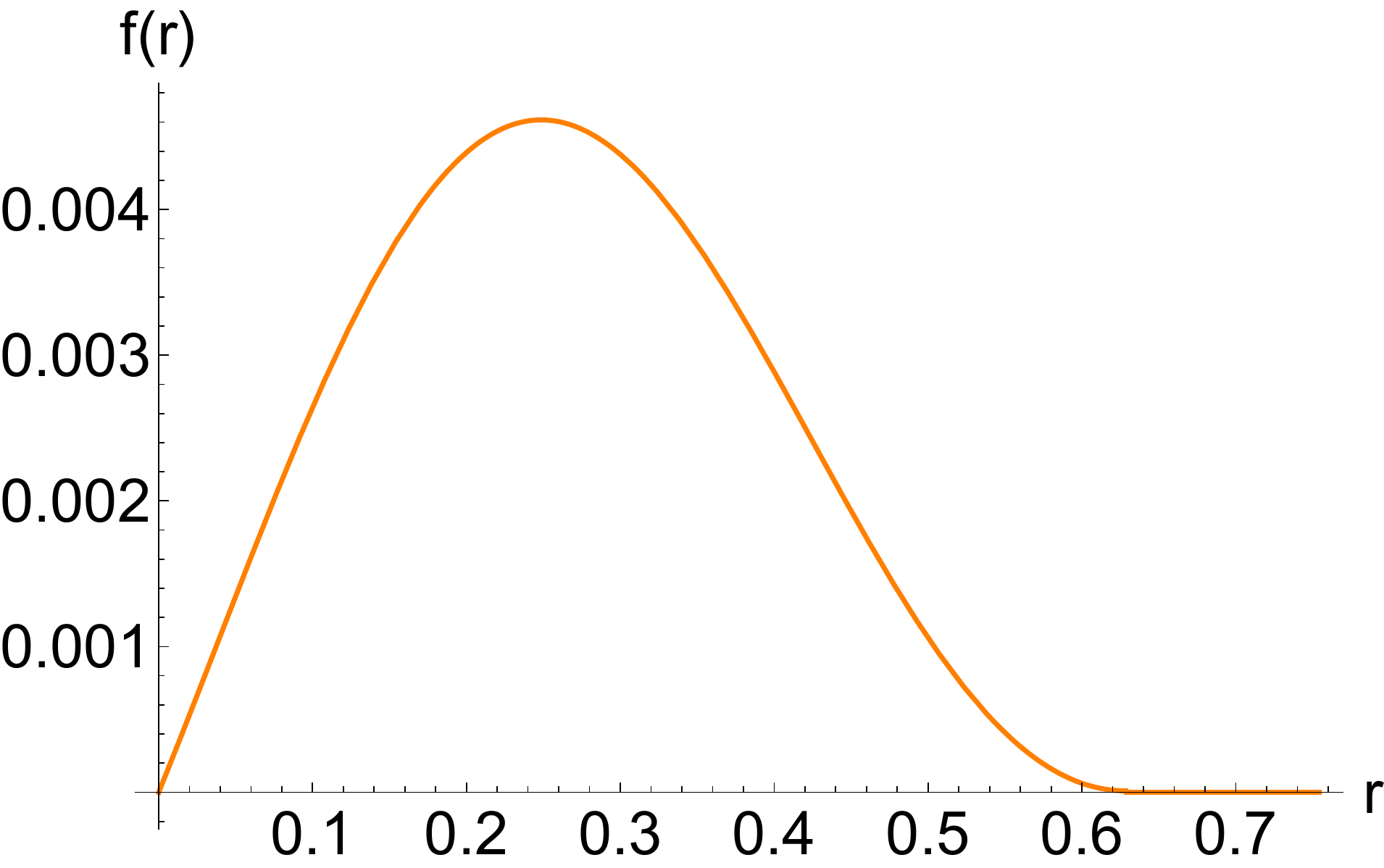}}
\subfigure[]{\includegraphics[width=0.3\textwidth,height=0.15\textwidth, angle =0]{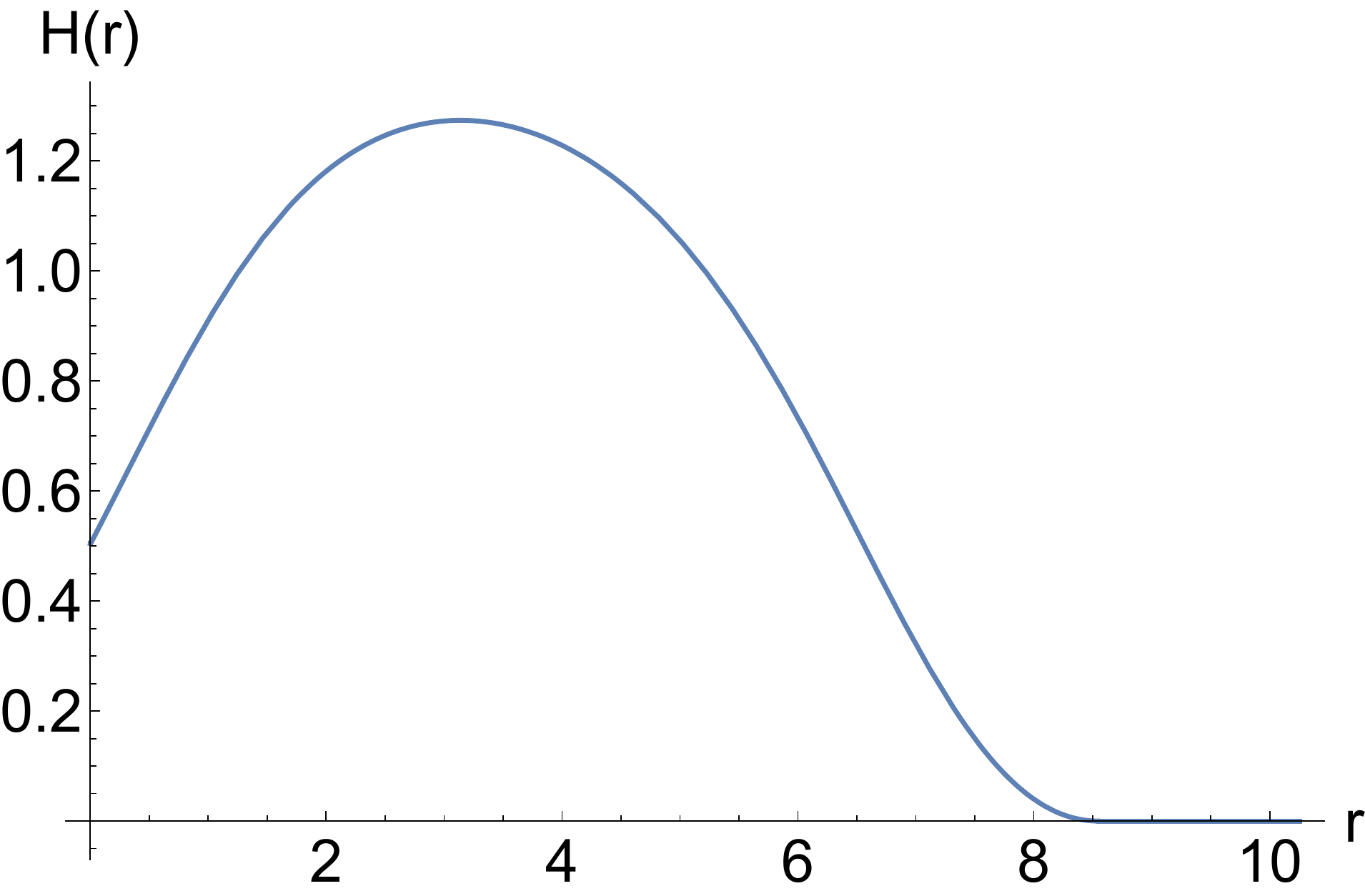}}
\subfigure[]{\includegraphics[width=0.3\textwidth,height=0.15\textwidth, angle =0]{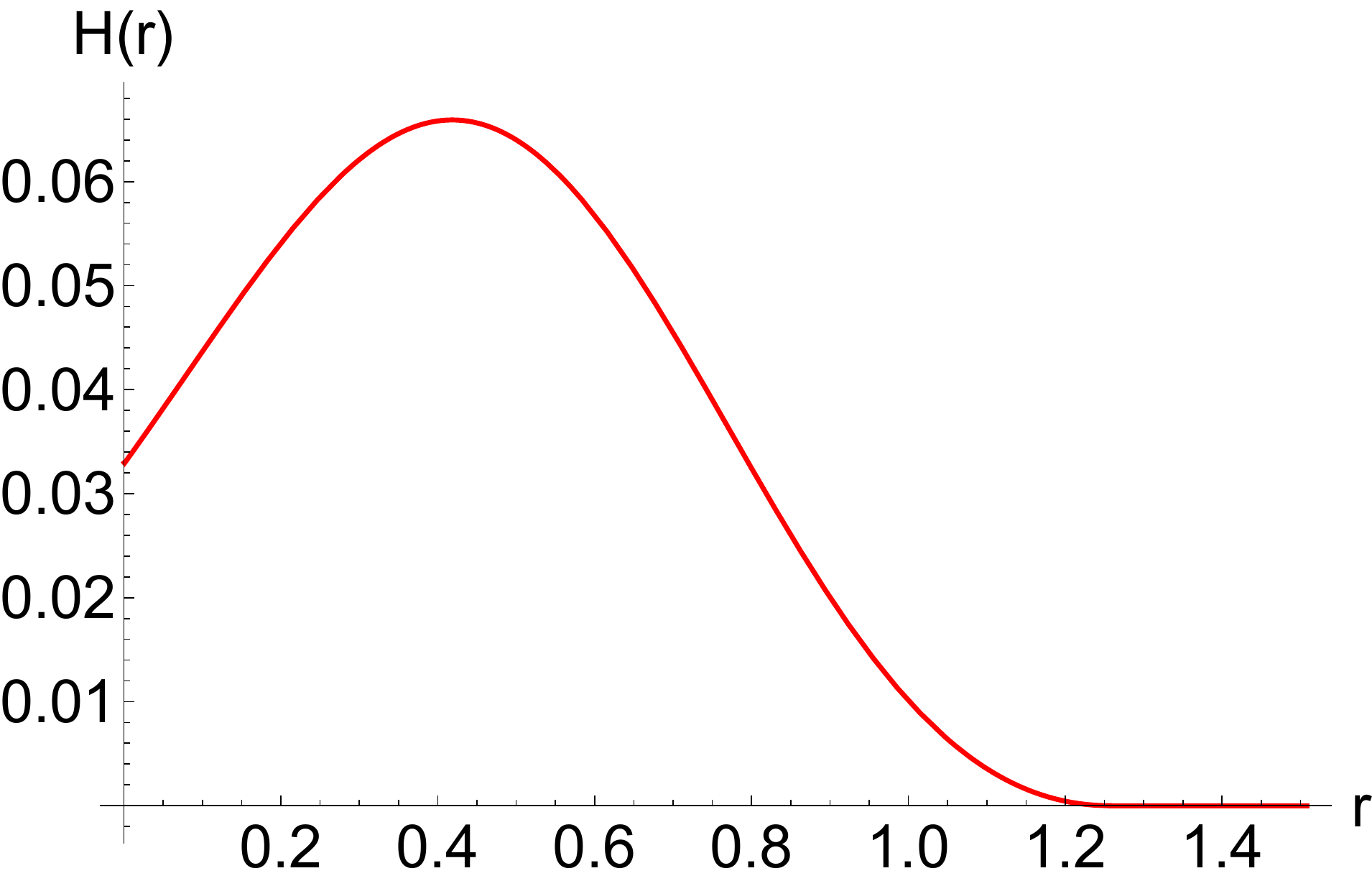}}
\subfigure[]{\includegraphics[width=0.3\textwidth,height=0.15\textwidth, angle =0]{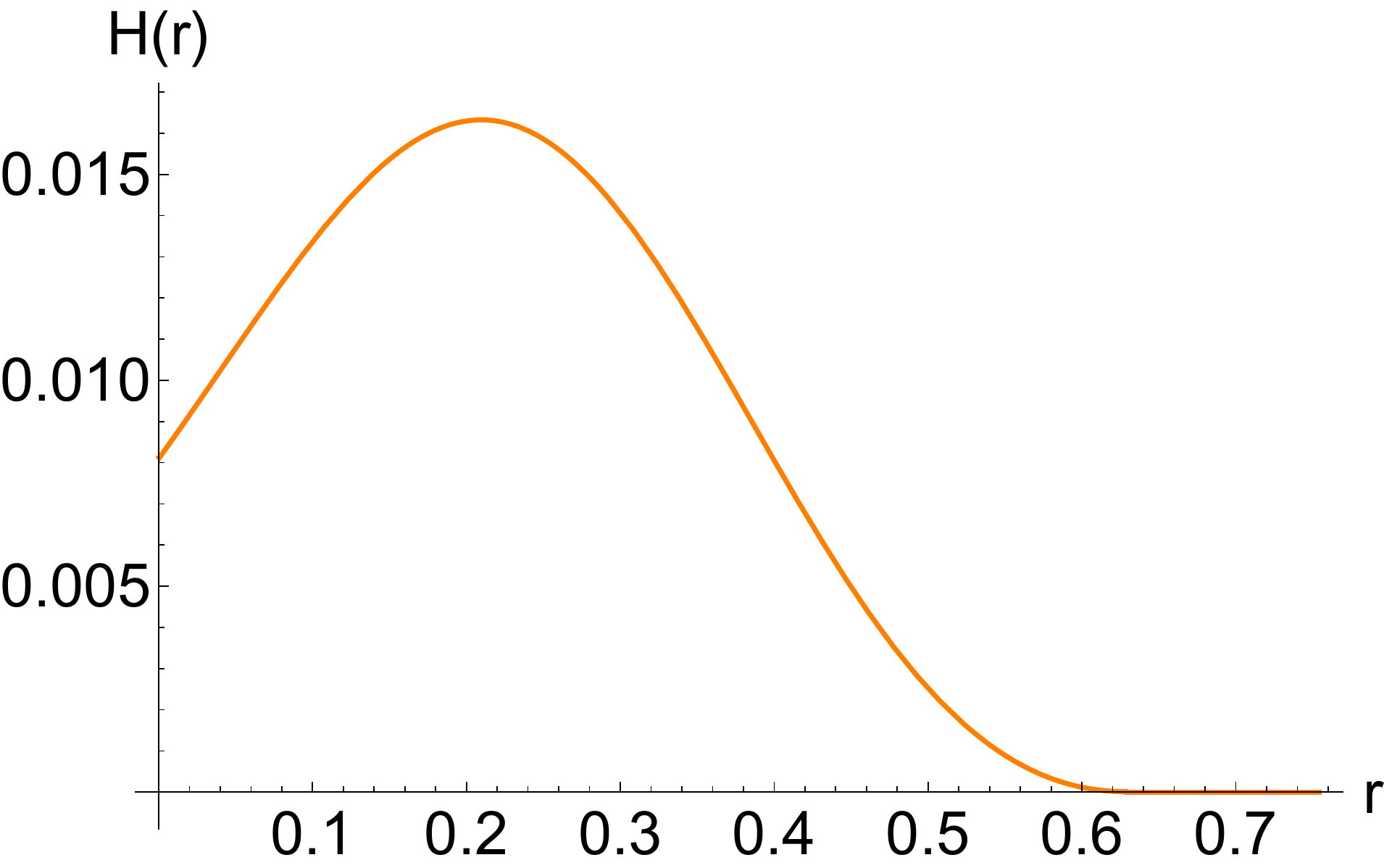}}
\caption{The radial function $f(r)$ and the energy density $H(r)$ for $l=1$ and (a,d) $\omega=1.0$, (b,e), $\omega=5.0$, (c,f) $\omega=10.0$.}\label{l=1}
\end{figure}

The profile functions $f(r)$ and the energy density plots are shown in Fig.\ref{l=1}. The fundamental difference between the cases $l=1$ and  $l=0$ case is a form of the solution at $r=0$. For $l=1$ the function $f(r)$ vanishes at the center whereas its first derivative $f'(r=0)$ is finite. The energy density  does not vanish at the center $r=0$, however, $H(0)$ is not a maximal value anymore. The maximum of $H(r)$ is reached at some finite distance from the center.

\begin{figure}[h!]
\centering
\subfigure[]{\includegraphics[width=0.3\textwidth,height=0.15\textwidth, angle =0]{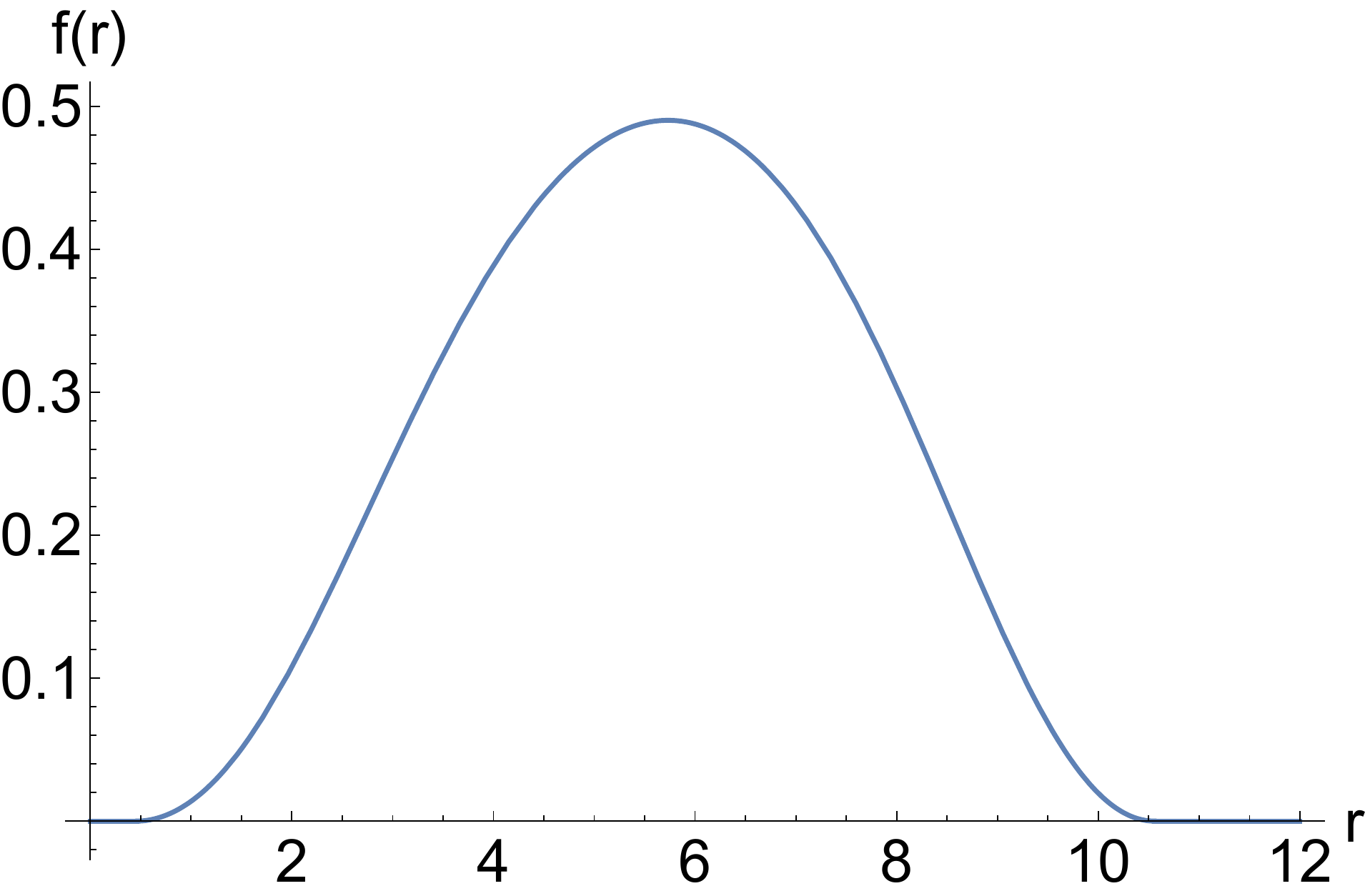}}
\subfigure[]{\includegraphics[width=0.3\textwidth,height=0.15\textwidth, angle =0]{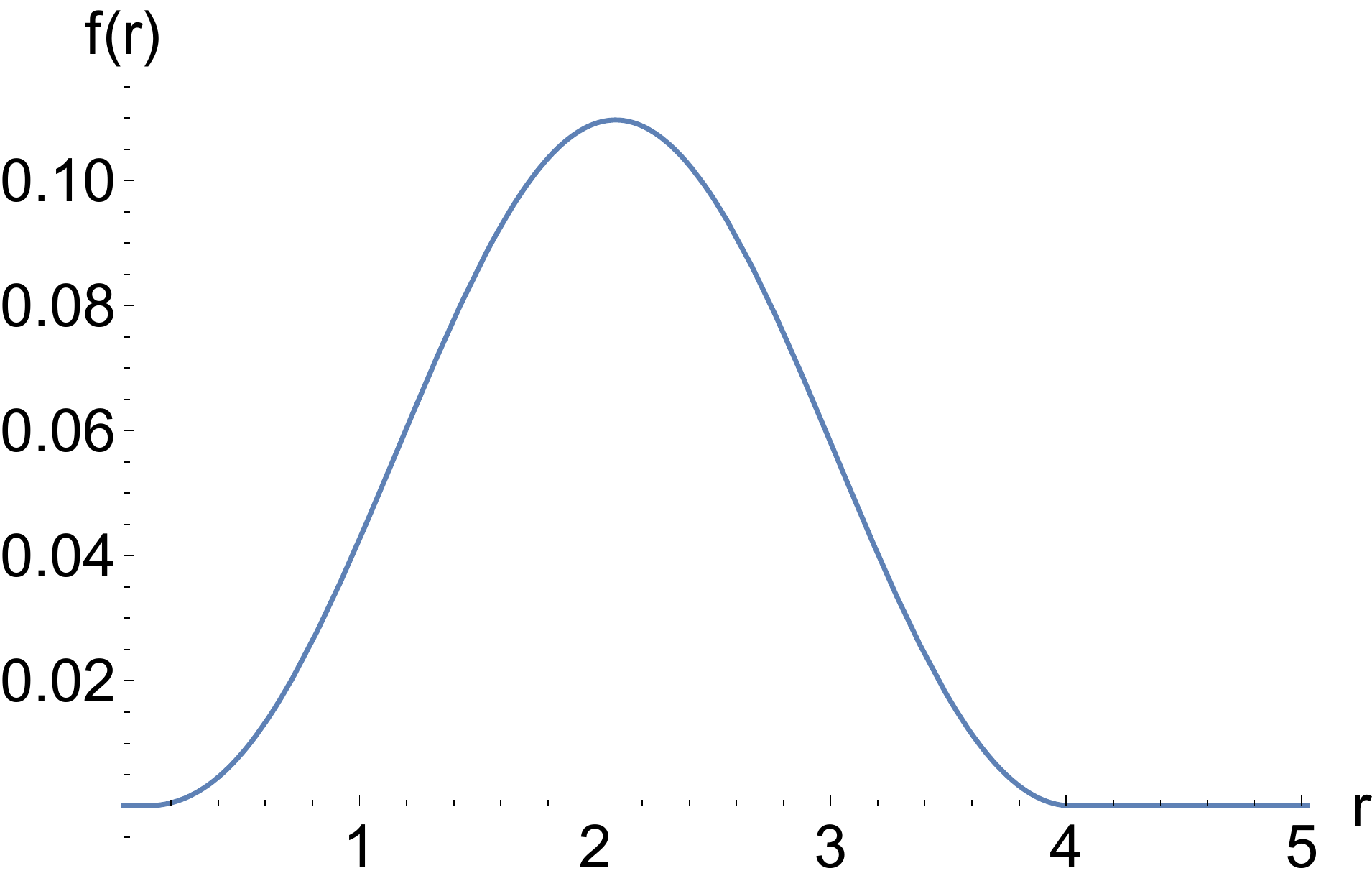}}
\subfigure[]{\includegraphics[width=0.3\textwidth,height=0.15\textwidth, angle =0]{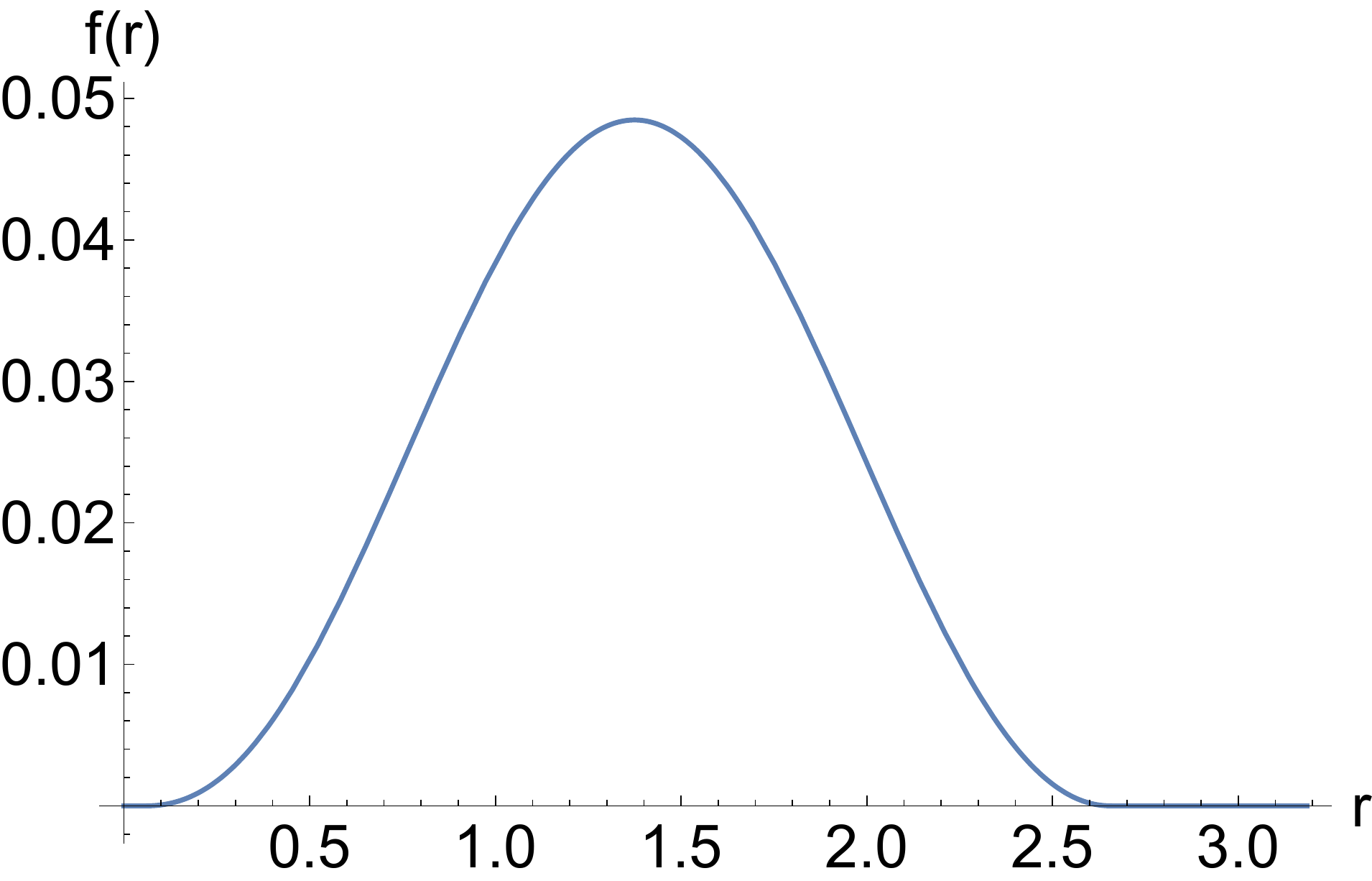}}
\subfigure[]{\includegraphics[width=0.3\textwidth,height=0.15\textwidth, angle =0]{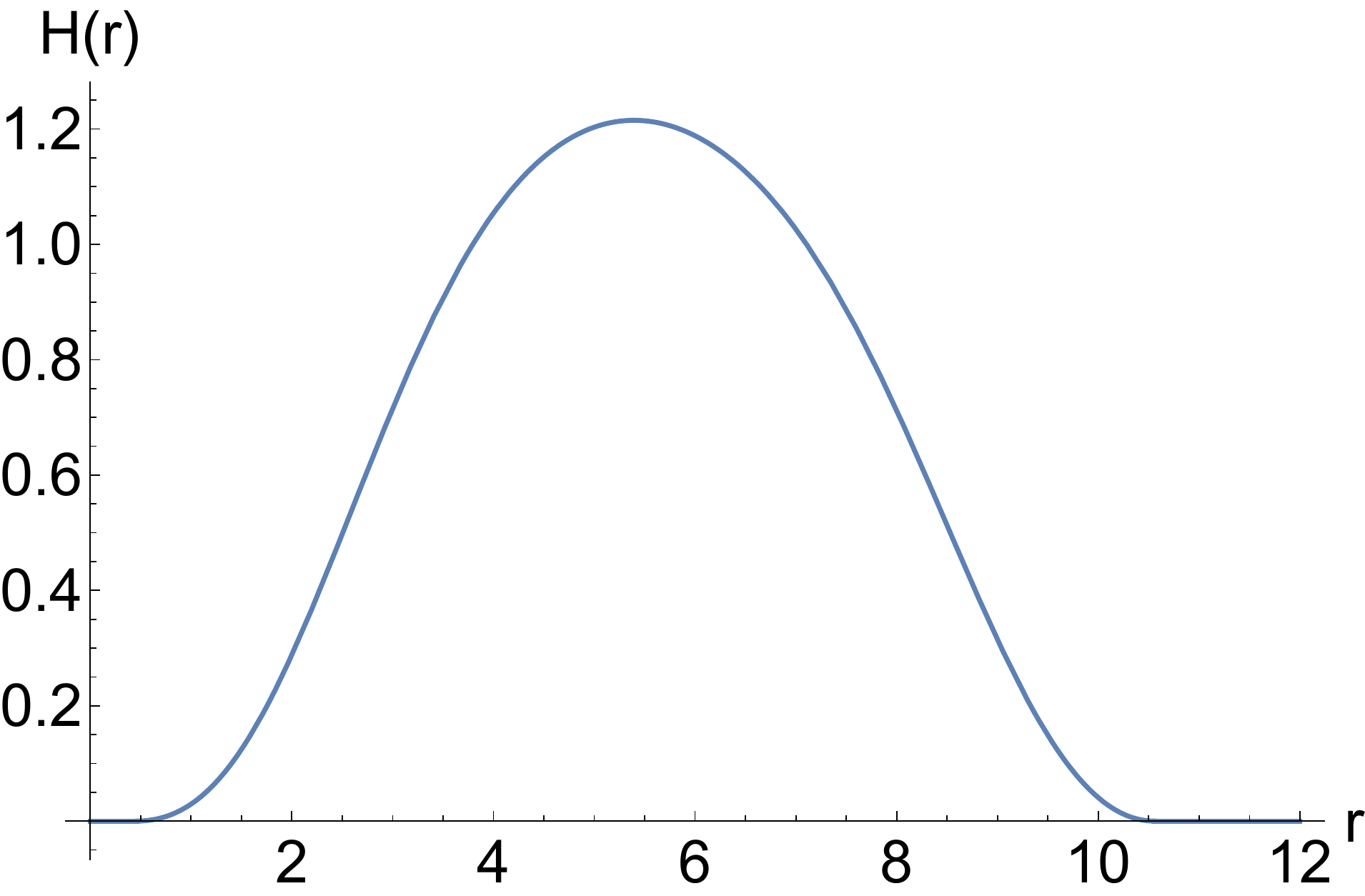}}
\subfigure[]{\includegraphics[width=0.3\textwidth,height=0.15\textwidth, angle =0]{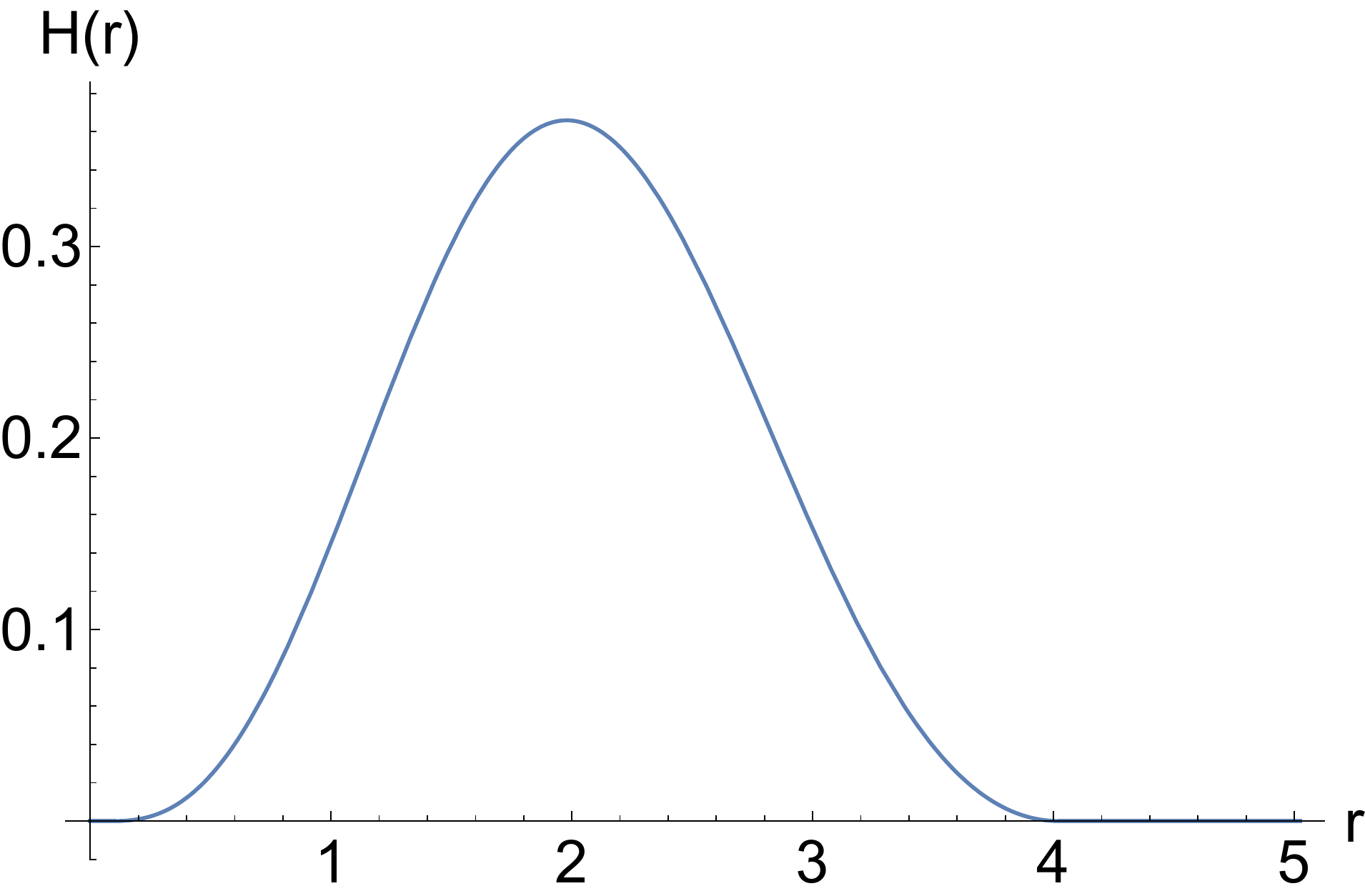}}
\subfigure[]{\includegraphics[width=0.3\textwidth,height=0.15\textwidth, angle =0]{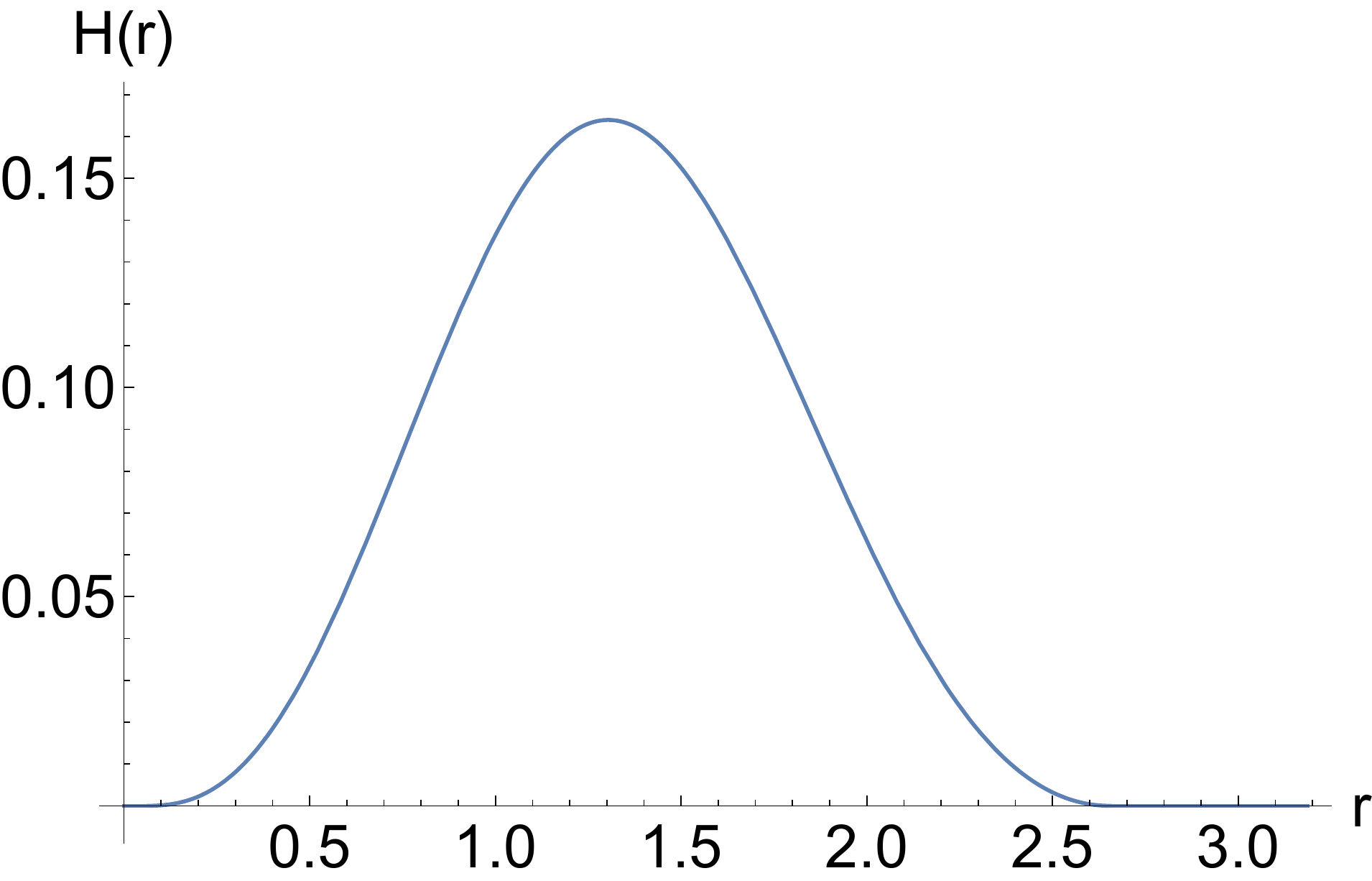}}
\caption{The radial function $f(r)$ and the energy density $H(r)$ for $l=2$ and (a,d) $\omega=1.0$, (b,e), $\omega=2.0$, (c,f) $\omega=3.0$.}\label{l=2}
\end{figure}

\begin{figure}[h!]
\centering
\subfigure[]{\includegraphics[width=0.3\textwidth,height=0.15\textwidth, angle =0]{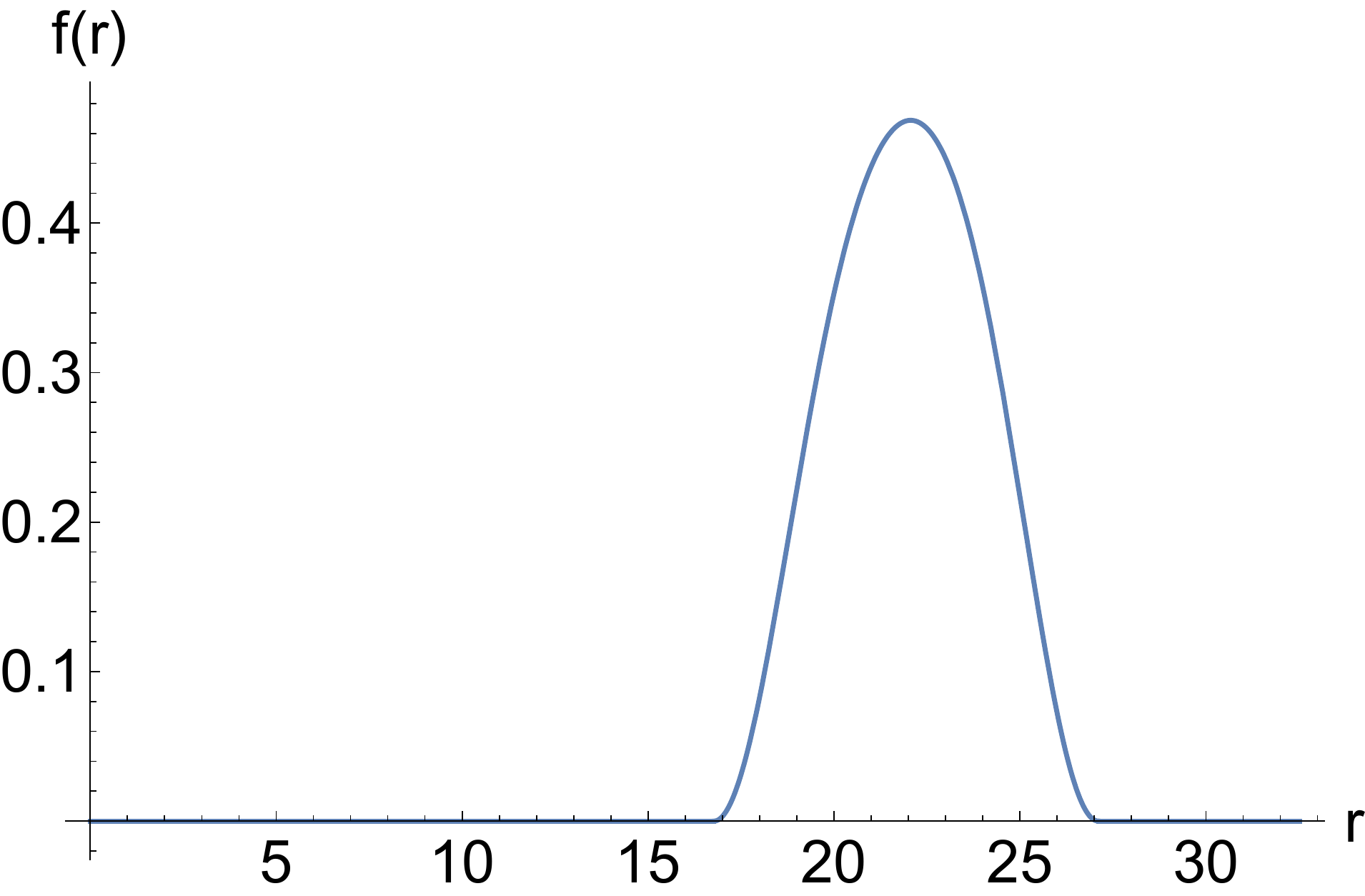}}
\subfigure[]{\includegraphics[width=0.3\textwidth,height=0.15\textwidth, angle =0]{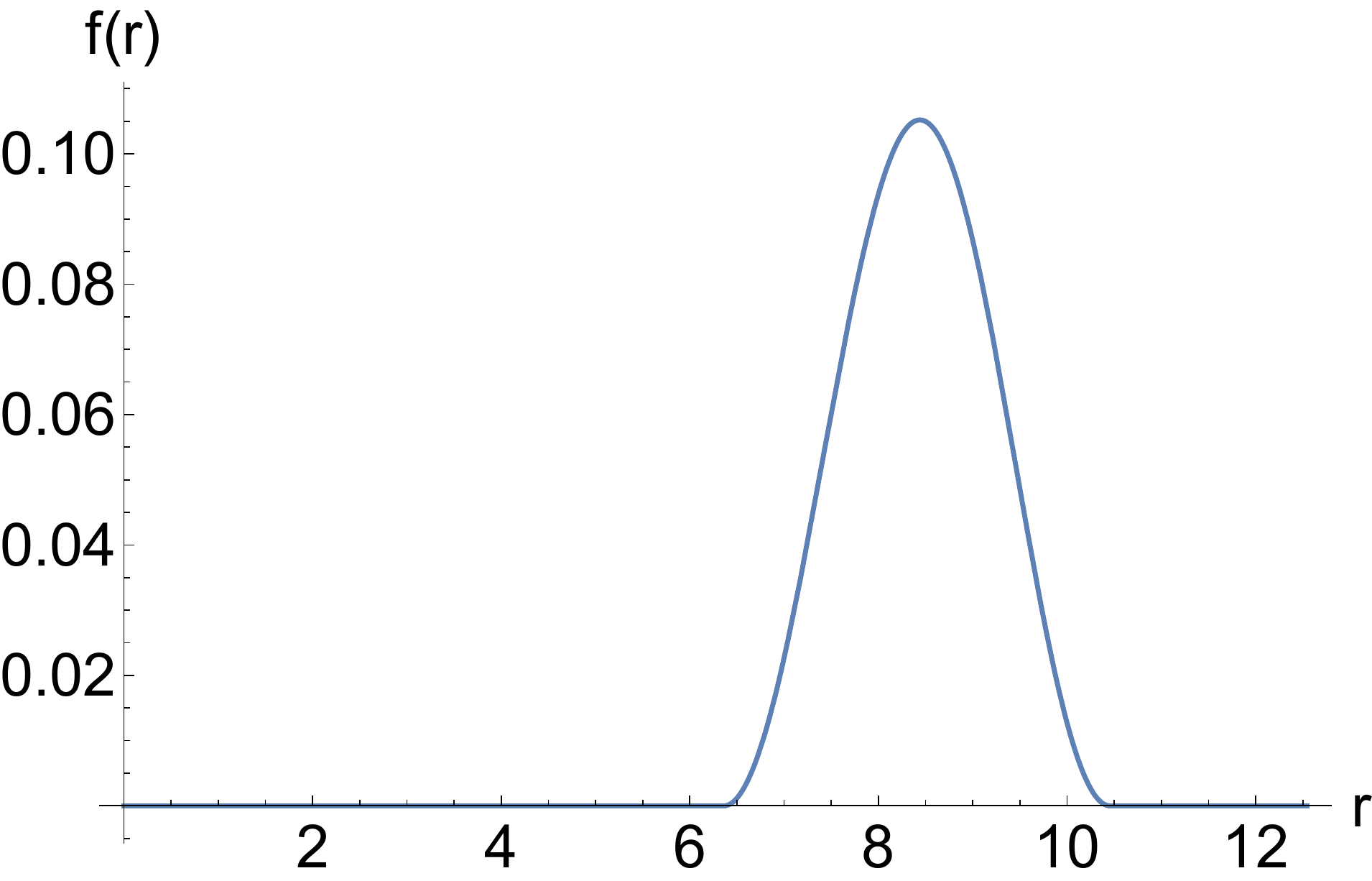}}
\subfigure[]{\includegraphics[width=0.3\textwidth,height=0.15\textwidth, angle =0]{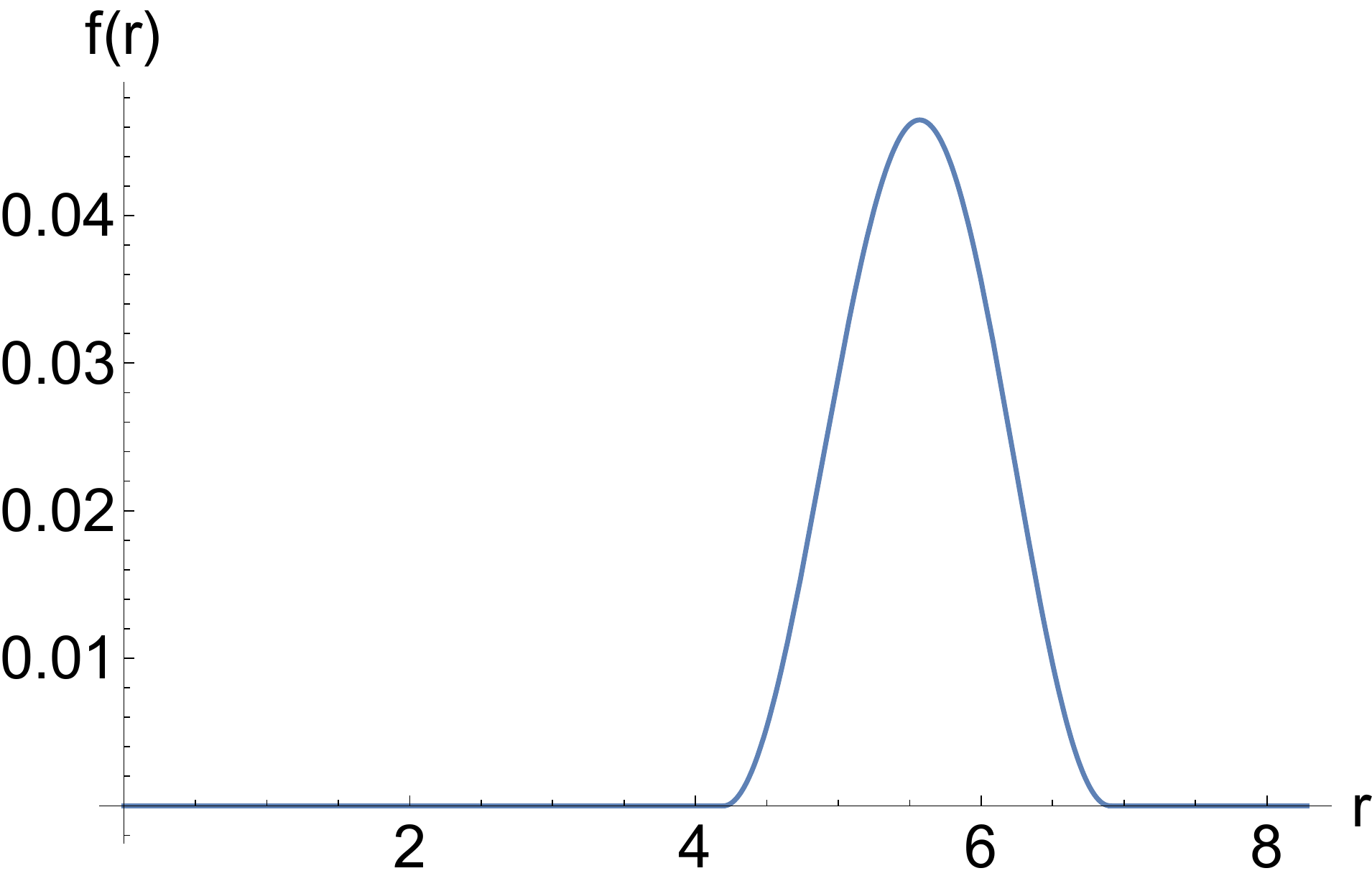}}
\caption{The radial function $f(r)$ for $l=10$ and (a) $\omega=1.0$, (b) $\omega=2.0$, (c) $\omega=3.0$.}\label{l=10}
\end{figure}

\begin{figure}[h!]
\centering
\subfigure[]{\includegraphics[width=0.45\textwidth,height=0.25\textwidth, angle =0]{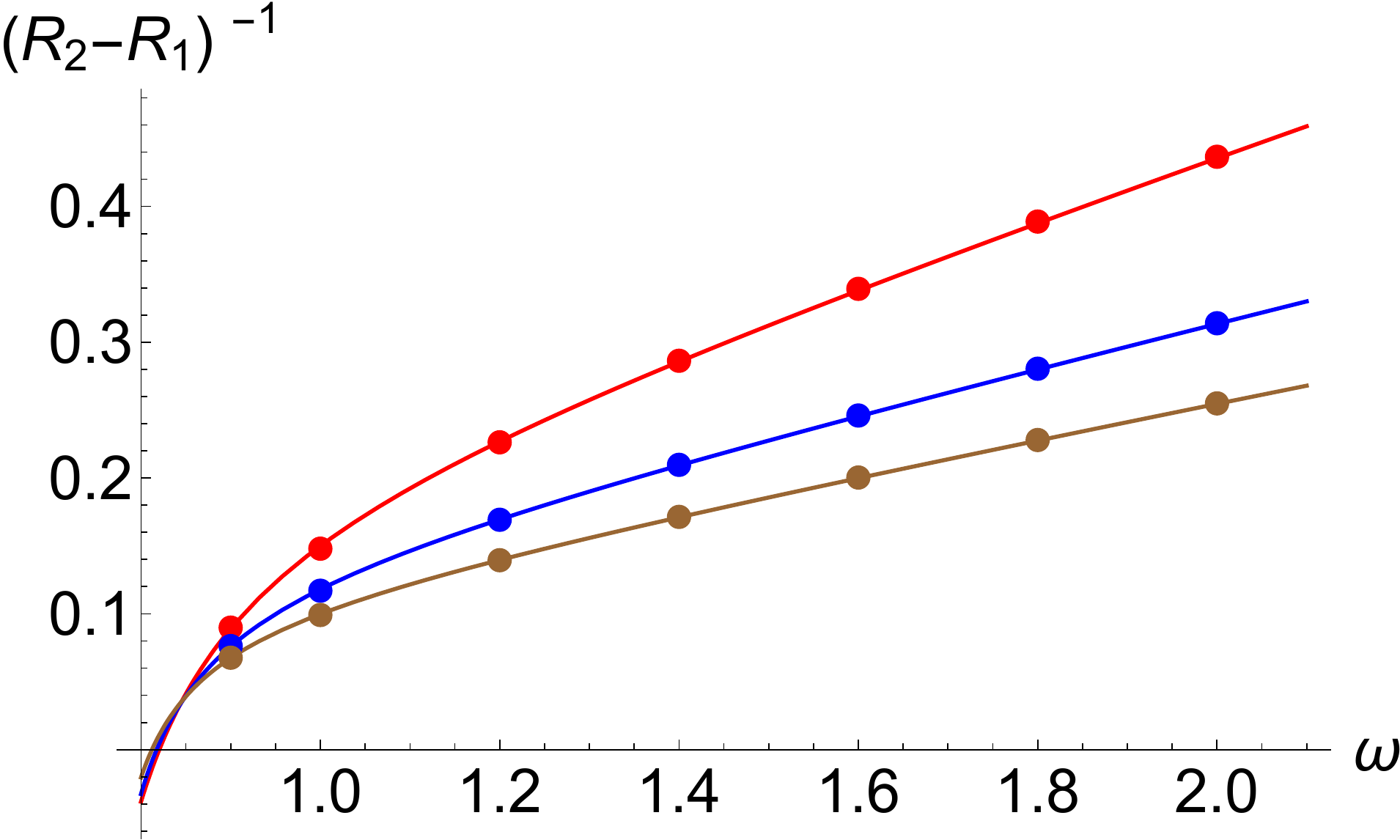}}
\subfigure[]{\includegraphics[width=0.45\textwidth,height=0.25\textwidth, angle =0]{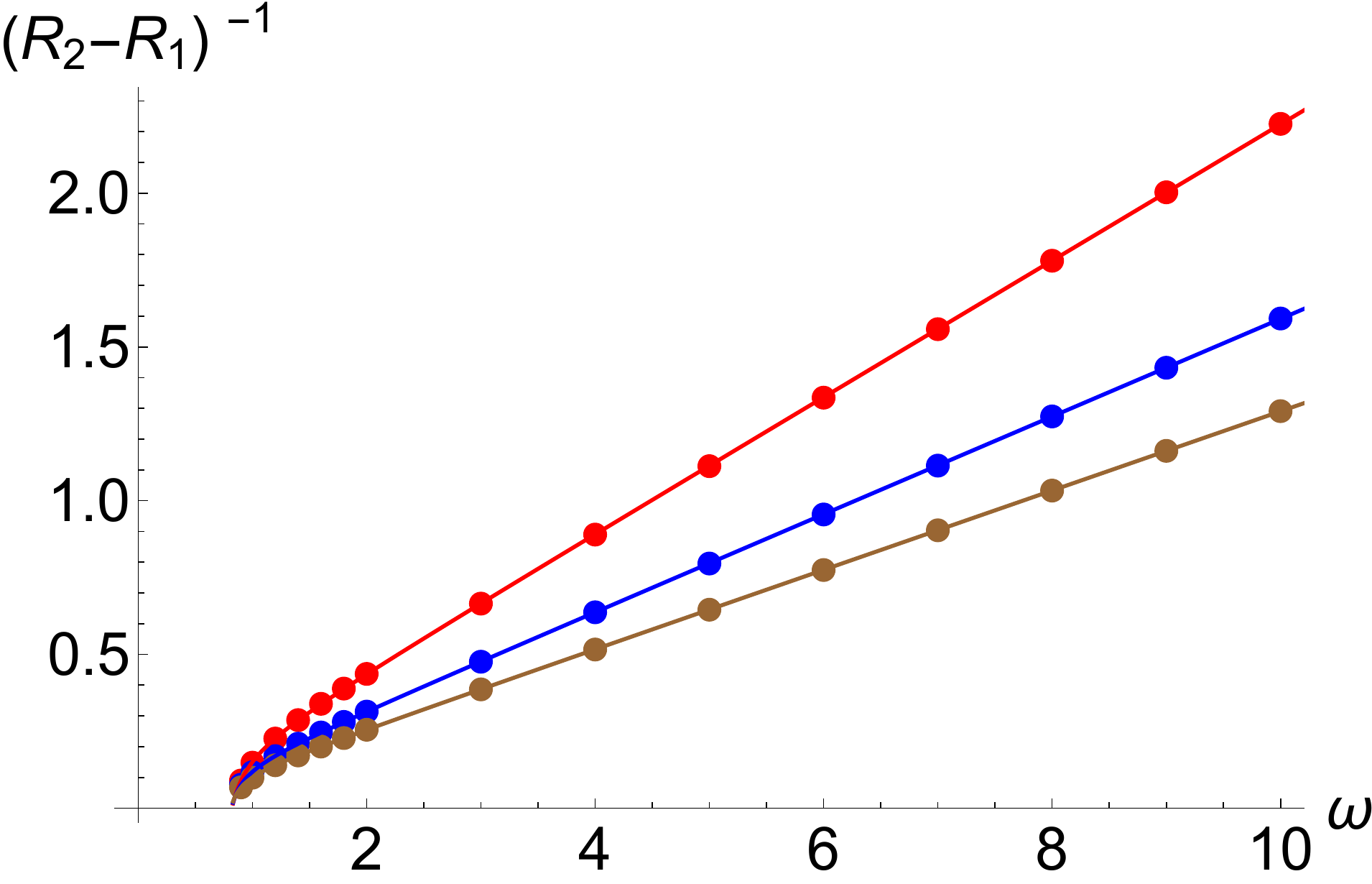}}
\caption{The inverse of the compacton size $\delta R:=R_2-R_1$ in dependence on $\omega$ for (from up to down) $l=0$, $l=1$ and $l=2$. The inner radius is $R_1\equiv 0$  for $l=0$ and $l=1$.}\label{compsize}
\end{figure}

Shooting parameter for $l\ge 2$ can be chosen as one of the compacton radii $R_1$ and $R_2$. We fine-tune a smaller radius $R_1$ in order to minimize the solution at bigger radius $f(\bar R)\rightarrow 0$ for $\bar R\rightarrow R_2$, where $\bar R$ is a solution of the equation $f'(\bar R)=0$. We interrupt the loop when accuracy $10^{-6}$ is reached. The function $f(r)$ is bell-shaped so each $u_m$ is non-trivial in the region of space given by spherical shell limited by the internal $R_1$ and the external $R_2$ radius.  The numerical values of compacton radii grows with the model parameter $l$. It can be easily seen comparing Fig.\ref{l=2} and Fig.\ref{l=10}. The compacton radii ($l\ge 2$) are decreasing functions of the parameter $\omega$. A very similar behaviour can be observed for $l=0,1$. In Fig.\ref{compsize} we plot the compacton size $\delta R:=R_2-R_1$ in dependence on the parameter $\omega$. Clearly, $R_1\equiv 0$ for $l=0,1$. For better transparency we plot a function $(\delta R)^{-1}(\omega)$. The function $(\delta R)^{-1}$ has linear asymptotic behaviour for $\omega\gg 1$. For small values $\omega$ the function $(\delta R)^{-1}(\omega)$ is not a linear function any longer, see Fig.\ref{compsize} (a). 
One of the simplest functions that can be fitted to numerical data is a rational function 
\be
(\delta R)^{-1}(\omega)=\frac{a\,\omega^2+b\,\omega+c}{d\,\omega+e}.\label{fit1}
\ee
Coefficients of fitted curves are presented in TABLE \ref{tab1}. 
\begin{table}[h!]
\centering
\label{my-label}
\begin{tabular}{|l||c|c|c|c|c|}
\hline
    & $a$ & $b$ & $c$ & $d$ & $e$  \\ \hline\hline
$l=0$ & $2223.78$  & $-1388.73$  & $-357.34$  &  $10044.60$ &   $-6859.62$   \\ \hline
$l=1$ & $53.86$  & $-36.61$  & $53.86$  & $39.62$  &   $-244.81$   \\ \hline
$l=2$ &  $169.46 $ &  $-118.27$ & $-15.73$  & $1316.28$  &   $-960.93$   \\ \hline
\end{tabular}\caption{The fit coefficients of \eqref{fit1}.}\label{tab1}
\end{table}

\noindent
Expression \eqref{fit1} has following asymptotic form for $\omega\rightarrow\infty$
\be
(\delta R)^{-1}(\omega)=\frac{a}{d}\,\omega +\frac{bd-ae}{d^2}+{\cal O}(\omega^{-1})
\ee
We denote coefficients $A_1:=\frac{a}{d}$ and $B_1:=\frac{bd-ae}{d^2}$. Their numerical values are presented in TABELE \ref{tab1a}.
\begin{table}[h!]
\centering
\label{my-label}
\begin{tabular}{|l||c|c||c|c|}
\hline
    & $A_1$ & $B_1$ & $A_2$ & $B_2$   \\ \hline\hline
$\;l=0\;$ & $\;0.221\;$  & $\;0.012\;$  & $\;0.369\;$  &  $\;0.014\;$    \\ \hline
$\;l=1\;$ & $0.158$  & $0.006$  & $0.287$  & $0.004$     \\ \hline
$\;l=2\;$ &  $0.128$ &  $0.004$ & $0.243$  & $0.005$     \\ \hline
\end{tabular}\caption{Coefficients of oblique asymptotes $\delta R^{-1}=A_1\omega+B_1$ and $E^{-1/5}=A_2\omega+B_2$.}\label{tab1a}
\end{table}

\begin{figure}[h!]
\centering
\subfigure[]{\includegraphics[width=0.45\textwidth,height=0.25\textwidth, angle =0]{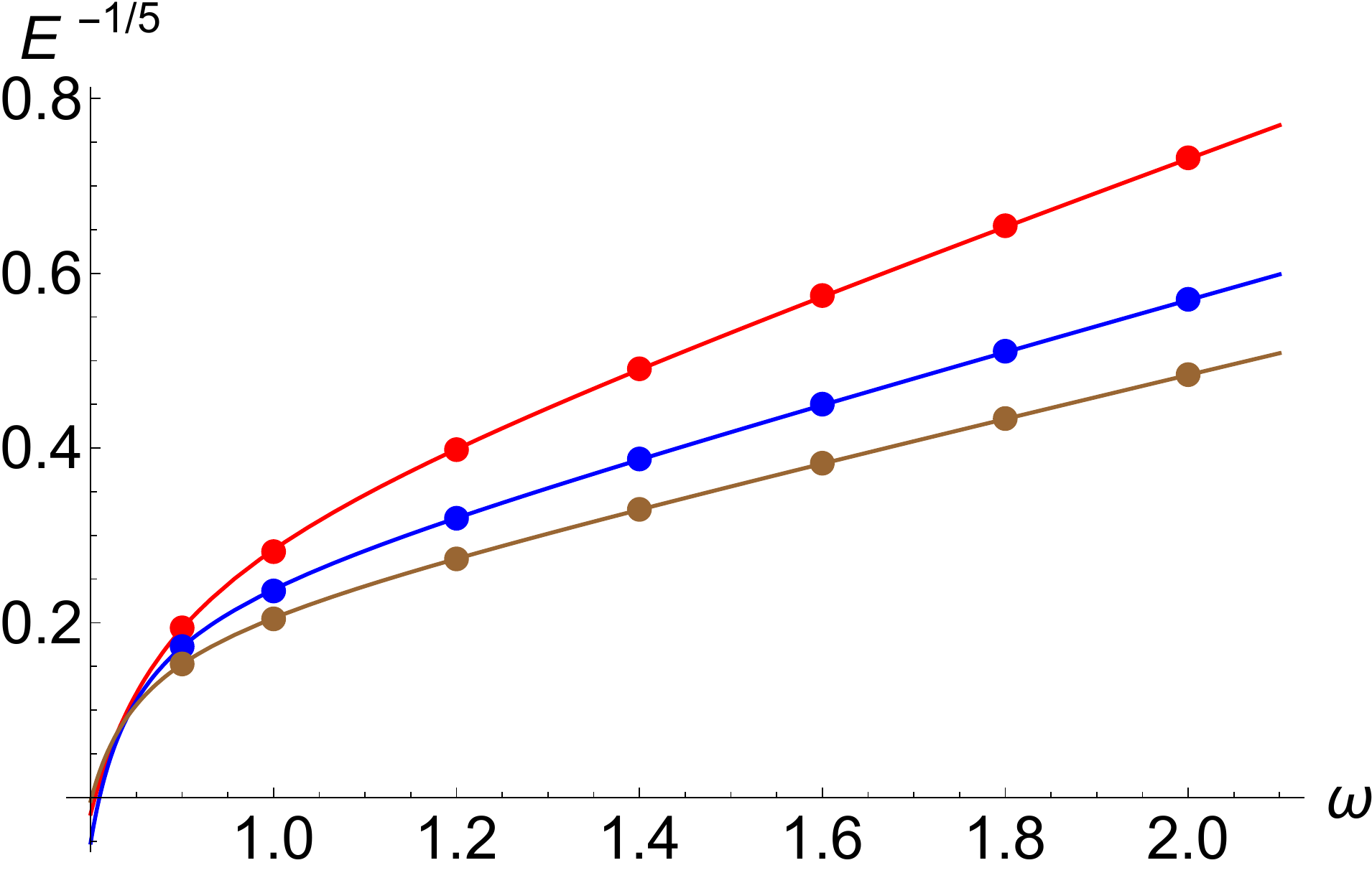}}
\subfigure[]{\includegraphics[width=0.45\textwidth,height=0.25\textwidth, angle =0]{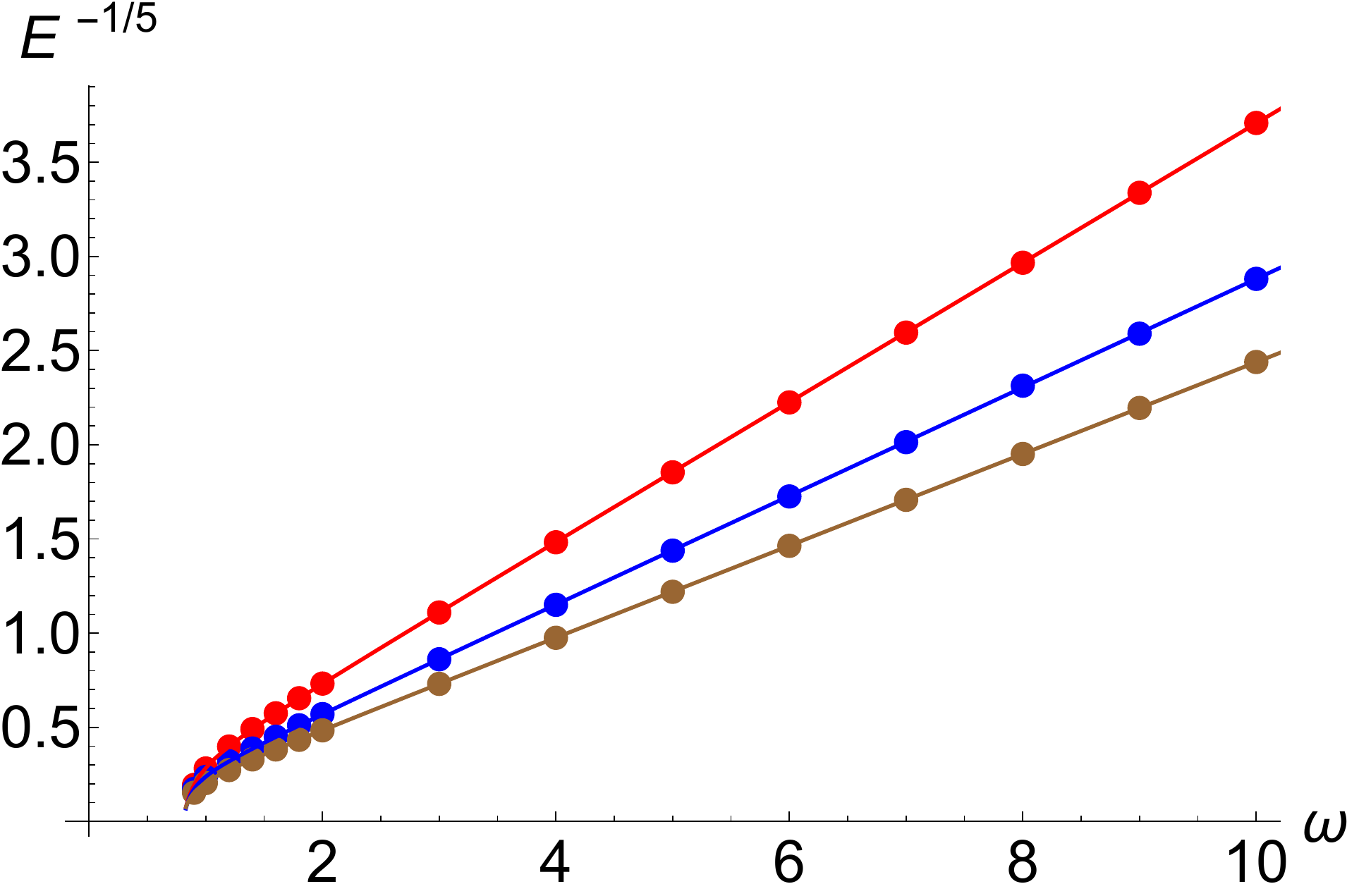}}
\caption{The compacton energy function in dependence on $\omega$ for (from up to down) $l=0$, $l=1$ and $l=2$.}\label{energy}
\end{figure}

Very similar analysis  can be performed for dimensionless energy of the compacton.  We observe that expression $E^{-1/5}$ is a linear function of $\omega$ for $\omega\gg 1$. The plot of this function in shown in Fig.\ref{energy}. Deviation from linear behaviour is observed for small values of $\omega$ . The curves that represent the fits are given by  rational functions $E^{-1/5}(\omega)=(\tilde a\,\omega^2+\tilde b\,\omega+\tilde c)/(\tilde d\,\omega+\tilde e)$. We shall not present the numerical  values of coefficients, instead,  we give in TABELE \ref{tab1a} the list of coefficients of asymptotic expression $E^{-1/5}=A_2\,\omega+B_2$ for $\tilde \mu=1$,  $\omega\gg 1$.

\begin{figure}[h!]
\centering
\subfigure[]{\includegraphics[width=0.45\textwidth,height=0.25\textwidth, angle =0]{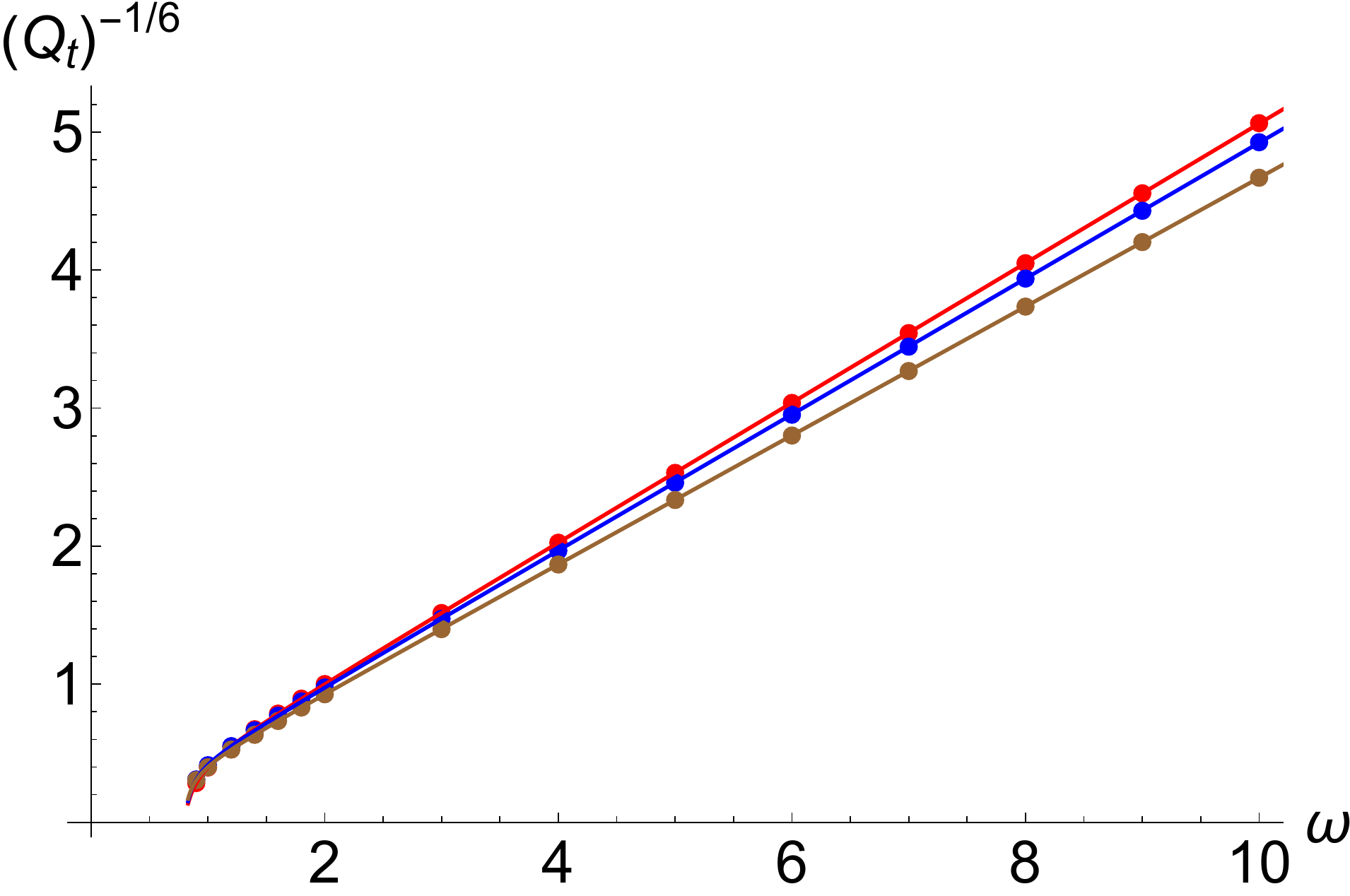}}
\subfigure[]{\includegraphics[width=0.45\textwidth,height=0.25\textwidth, angle =0]{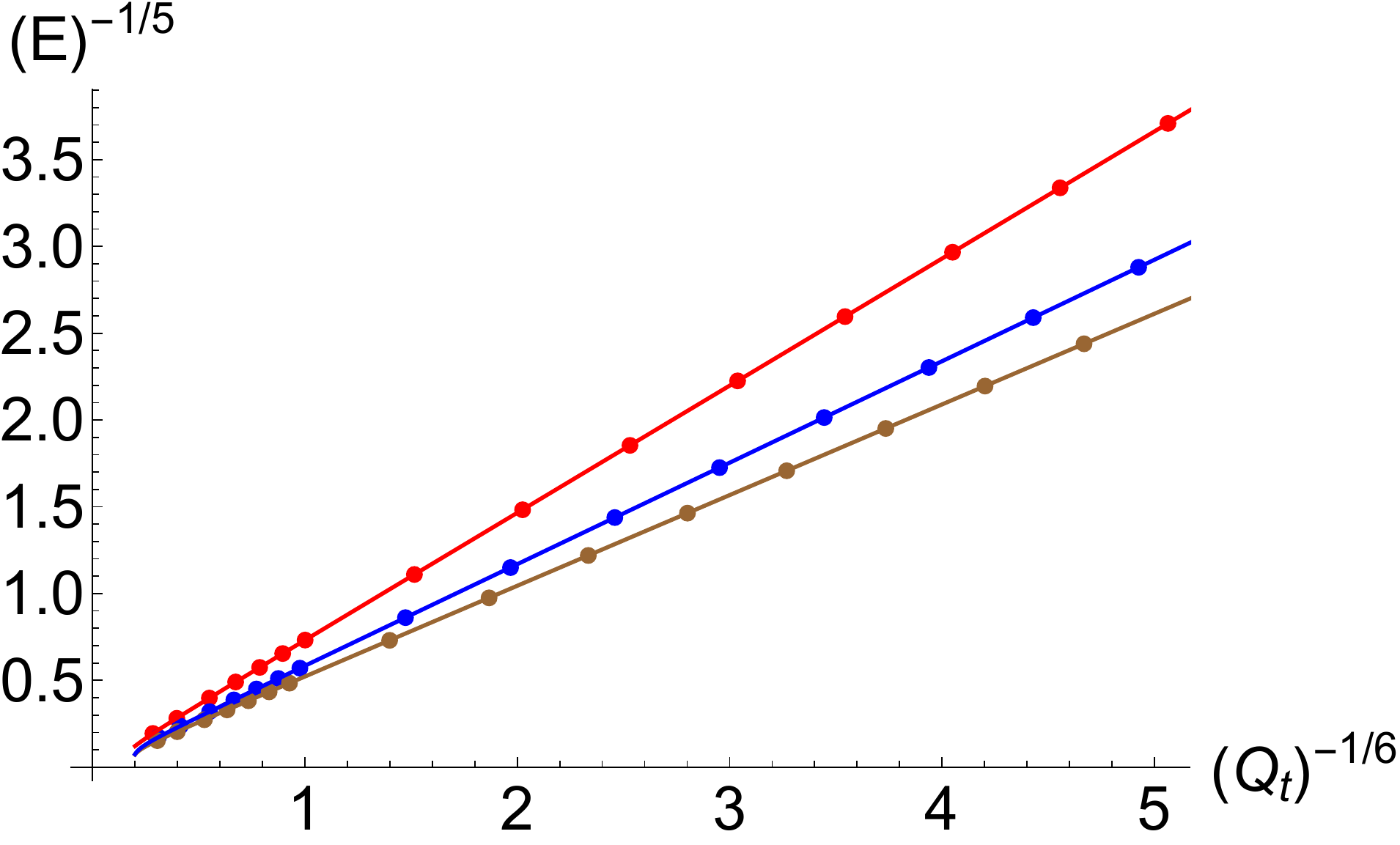}}
\caption{(a) Noether charge $Q_t$ for (from up to down) $l=0$, $l=1$ and $l=2$ in dependence on $\omega$. (b)  A relation between Noether charges and the energy of the solution  for (from up to down) $l=0$, $l=1$ and $l=2$.}\label{Qw}
\end{figure}

Another important point is an analysis of the Noether charges and theirs relation with the energy of the solution. 
The plot of these charges is presented in Fig.\ref{Qw} (a). We observe that for $\omega\gg 1$ the charges behaves as $Q_t\propto\omega^{-6}$. We shall omit the index $m$ because the Noether charges do not depend on it. In Fig.\ref{Qw} (b) we plot relation energy-charge for Q-balls $l=0,1$ and Q-shells $l=2$. 
The leading behaviour of the function $Q^{-1/6}_{t}$ in the limit $\omega\gg 1$ is given by $Q^{-1/6}_{t}=A_3\, \omega +B_3$. Numerical values of coefficients $A_3$ and $B_3$ are presented in TABELE \ref{tab2}. We also present coefficients $A_4$ and $B_4$ which are given as coefficients in expression $E^{-1/5}=A_4Q_t^{-1/6}+B_4$. 
\begin{table}[h!]
\centering
\label{my-label}
\begin{tabular}{|l||c|c||c|c|}
\hline
    & $A_3$ & $B_3$ & $A_4$ & $B_4$   \\ \hline\hline
$\;l=0\;$ & $\;0.504\;$  & $\;0.017\;$  & $\;0.731\;$  &  $\;0.003\;$    \\ \hline
$\;l=1\;$ & $0.491$  & $0.008$  & $0.584$  & $0.001$     \\ \hline
$\;l=2\;$ &  $0.466$ &  $0.009$ & $0.522$  & $0.001$     \\ \hline
\end{tabular}\caption{Coefficients of oblique asymptotes $Q_t^{-1/6}=A_3\omega+B_3$ and $E^{-1/5}=A_4Q_t^{-1/6}+B_4$}\label{tab2}
\end{table}
One can conclude from Fig. \ref{Qw}(b) that the relation between the energy $E^{-1/5}$ and the Noether charges $Q_t^{-1/6}$ is linear with a very good accuracy, even though in the region of small $\omega$. It means that the energy of compactons behaves as $E\propto Q_t^{\frac{5}{6}}$ in whole range of $\omega$. The value of the power suggest that splitting a single Q-ball solution into two Q-balls is not energetically favorable because $E(Q_1+Q_2)<E(Q_1)+E(Q_2)$. This argument is usually presented in discussion of  stability of Q-ball solutions \cite{al1}.

\begin{figure}[h!]
\centering
\subfigure[]{\includegraphics[width=0.4\textwidth,height=0.25\textwidth, angle =0]{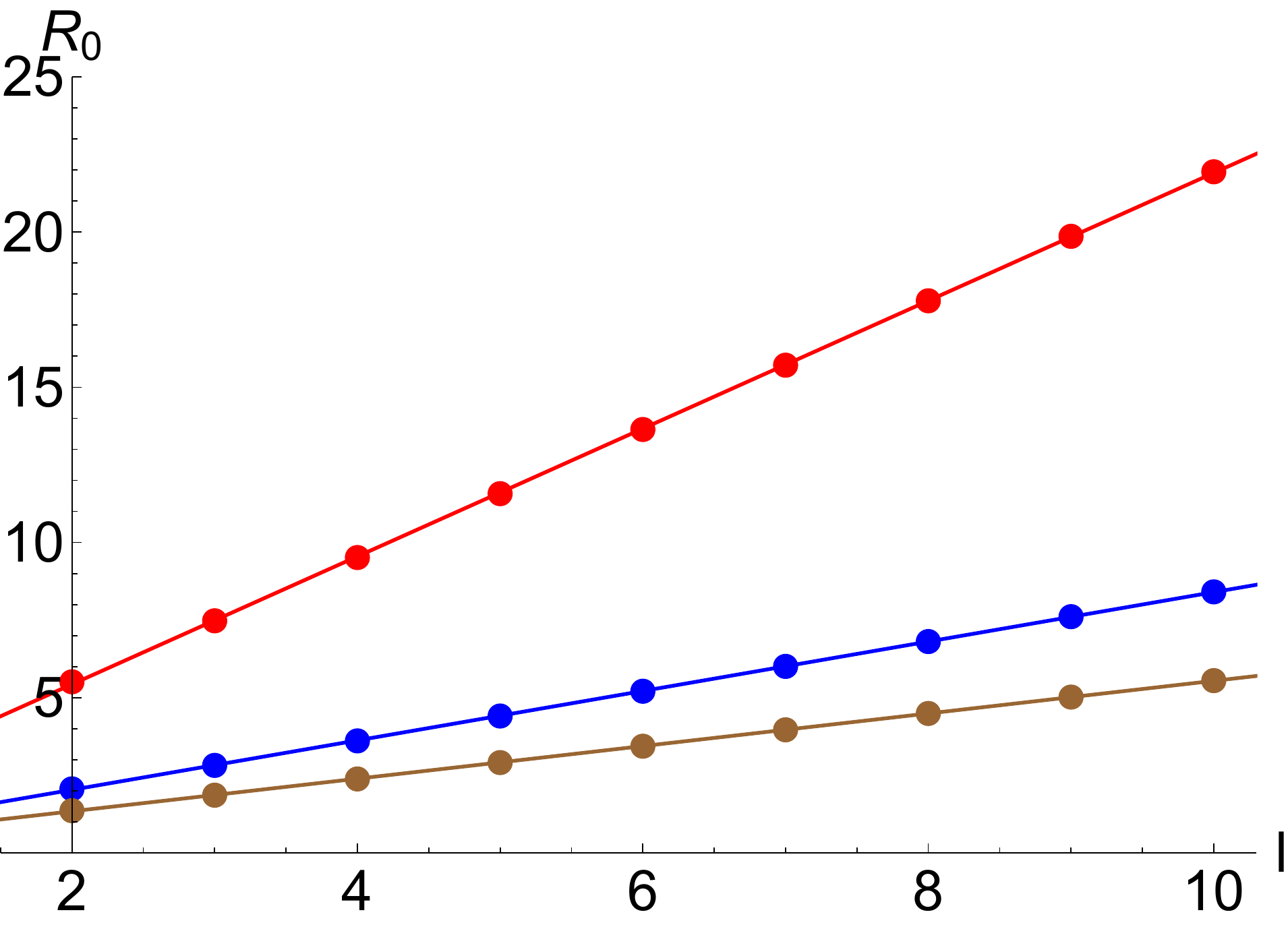}}
\subfigure[]{\includegraphics[width=0.4\textwidth,height=0.25\textwidth, angle =0]{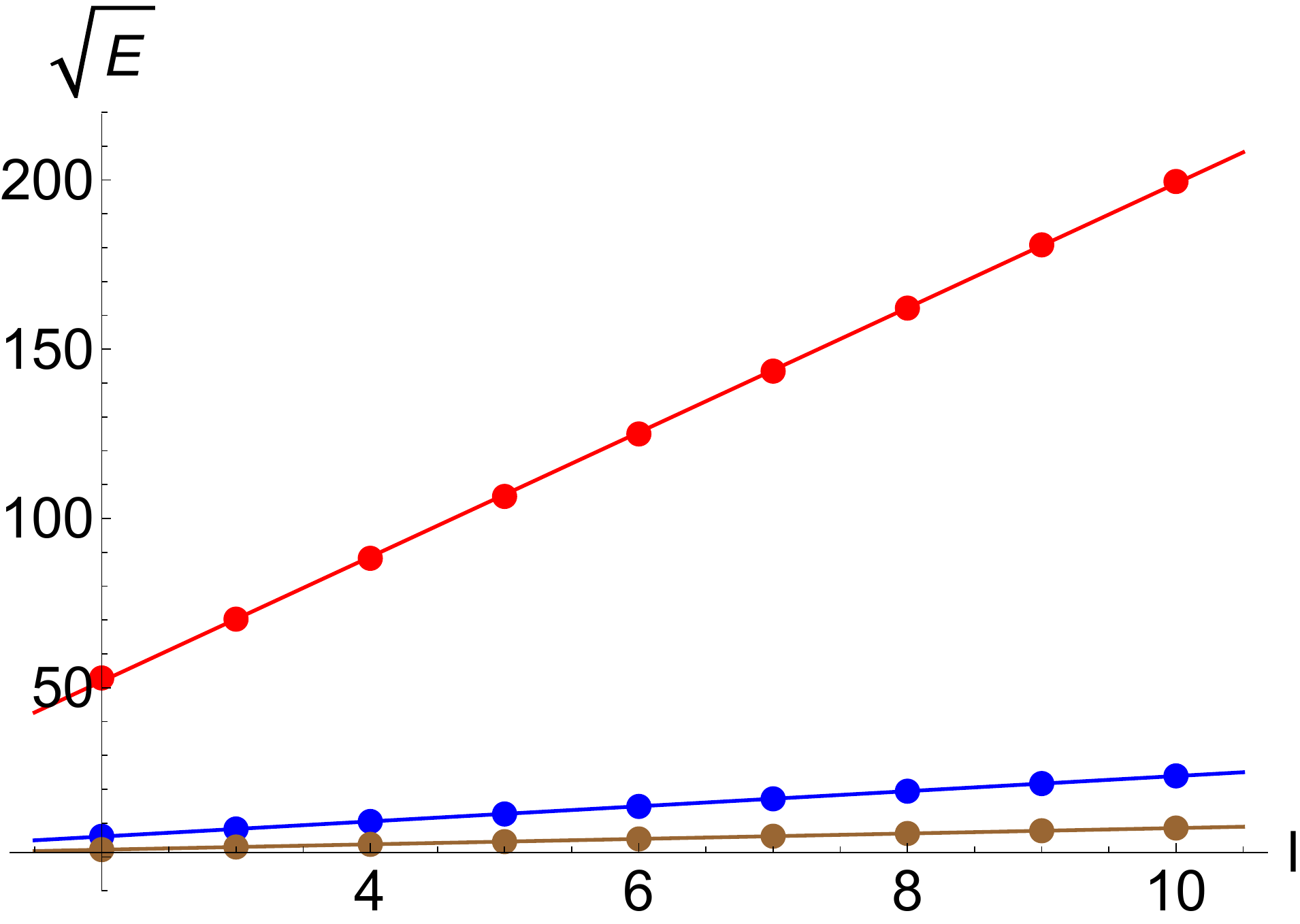}}
\caption{(a) The  medium radius $R_0=\frac{1}{2}(R_1+R_2)$ of compact shells in dependence on $l=2,\ldots,10$. From up to down:   $\omega=1.0$,  $\omega=2.0$ and $\omega=3.0$.  (b) The compacton energy sqare root  in dependence on $l=2,\ldots,10$.}\label{energy1}
\end{figure}

Finally, we plot the medium radius of compact shells $R_0:=\frac{1}{2}(R_1+R_2)$ in dependence on  $l$. Fig.\ref{energy1} (a) shows $R_0$ for $l=2,3,\ldots,10$ and for three different values of $\omega=1.0$, $\omega=2.0$ and $\omega=3.0$. The medium radius of compacton grows linearly with $l$. A linear fit gives
$R_0\approx1.31 + 2.06\,l$ for $\omega=1.0$, $R_0\approx 0.44 + 0.79\,l$ for $\omega=2.0$, $R_0\approx 0.29 + 0.52\,l$ for $\omega=3.0$. The medium radius $R_0$ decreases as $\omega$ grows. 

In Fig.\ref{energy1} (b) we show the square root of the energy of compactons in dependence on $l$. Linear fits are given by $\sqrt{E}\approx 15.13 + 18.38\,l$  for $\omega=1.0$,  $\sqrt{E}\approx 1.59 + 2.23\,l$ for $\omega=2.0$, $\sqrt{E}\approx 0.56 + 0.80\,l$ for $\omega=3.0$. Note, that linear regression is useful for extrapolation of functions $R_0(l)$ and $\sqrt{E(l)}$ to higher integers $l$ whereas interpolation to non-integer values is meaningless.


\section{The signum-Gordon limit}
According to our numerical results,  the size of the compactons, their energy and the Noether charges behave in the limit $\omega\rightarrow \infty$ as some powers of $\omega$. In this section we shall study this problem from analytic point of view. There are many exact results which can be obtained for a model which is a limit version of the $CP^N$ type model \eqref{lagr2}.

Numerical analysis shows that maximal value of solutions $f(r)$ and $|f'(r)|$ tend to zero as $\omega$ increases. Clearly, in this limit the model can be approximated by the complex signum-Gordon type model with equation of motion
\be
\partial^2 u_i+\frac{\mu^2}{8M^2}\frac{u_i}{\sqrt{u^{\dagger}\cdot u}}=0\label{sign-gordon-eq}
\ee
for $u_i\neq 0$ and $\partial^2 u_i=0$ for $u_i=0$.
To be more precise, the equation  \eqref{sign-gordon-eq} is the complex signum-Gordon equation of motion only for $l=0$ {\it i.e.} when the model possesses exactly one complex field $u$. For $l\ge 1$ the model is parametrized by $2l+1$ complex fields coupled via potential term. For this reason we call it the signum-Gordon {\it type} model. The solutions of the model described by \eqref{sign-gordon-eq} can be seen as limit solutions $|u_i|\ll 1$ of the $CP^N$ type model discussed above. In the further part of this secton we show that the proportionality relations $\delta R\propto \omega^{-1}$ and $E\propto \omega^{-5}$ are exact for the signum-Gordon type model. Next we shall discuss relation between  the energy and the Noether charges.

The equations of motion \eqref{sign-gordon-eq} for the ansatz \eqref{ansatz} is reduced to a single radial equation. In terms of new radial variable $x:=\omega\, r$ the radial equation takes the form
\be
\tilde f''(x)+\frac{1}{x}\tilde f'(x)+\left(1-\frac{l(l+1)}{x^2}\right)\tilde f(x)=\alpha ^2\, {\rm sgn}(\tilde f(x)),\label{besselnonhom}
\ee
where $\tilde f(x):=f(\frac{x}{\omega})\equiv f(r)$, $\tilde f'(x)\equiv\frac{d\tilde f}{dx}$, $\alpha ^2:=\frac{\tilde\mu^2}{8\omega^2}$ and $\tilde\mu^2:=\frac{\mu^2}{M^4}$ is a dimensionless  coupling constant defined in the same way as for the $CP^N$ type model. The energy density is given by expression
\be
H(r)=4\omega ^2\left[\left(\frac{d\tilde f}{dx}\right)^2+\left(1+\frac{l(l+1)}{x^2}\right)\tilde f^2\right]+\tilde\mu^2 \tilde f{\rm sgn}\,(\tilde f).\label{sign-gordon-ed}
\ee
The equation \eqref{besselnonhom} is a  spherical Bessel's equation, non-homogeneous for ${\rm sgn}(\tilde f)=1$ and homogeneous for ${\rm sgn}(\tilde f)=0$. The radial equation possesses exact solutions. The compact solutions consist of non-trivial solutions of the non-homogeneous  equation which are  matched  with the vacuum solution $\tilde f=0$.  In the case ${\rm sign\,}(\tilde f)=1$ the solution is a sum of general solution of the homogeneous equation and any particular solution $\tilde f_p(x)$ of the non-homogeneous equation {\it i.e.}
\be
\tilde f(x)={\cal A}\,j_l(x)+{\cal B}\,n_l(x)+\tilde f_p(x),\label{solgeral}
\ee
where ${\cal A}$ and ${\cal B}$ are free constants. The spherical Bessel functions $j_l(x)$ and the spherical Neumann functions $n_l(x)$ form  linearly independent solutions of the spherical Bessel's equation so their Wronskian is different from zero
\be
W(x)=j_l(x)n'_l(x)-j'_l(x)n_l(x)=\frac{1}{x^2}\neq 0.\label{totalsol}
\ee
The particular solution can be determined by the method of variation of parameters {\it i.e.} it is of the form
\be
\tilde f_p(x)=a(x)j_l(x)+b(x)n_l(x),\label{particularsol}
\ee
where $a(x)$ and $b(x)$ must be such that they satisfy equations $a'(x)j_l(x)+b'(x)n_l(x)=0$ and $a'(x)j'_l(x)+b'(x)n'_l(x)=\alpha^2$. They have solutions $a'(x)=-\frac{\alpha^2n_l(x)}{W(x)}$ and $b'(x)=\frac{\alpha^2\,j_l(x)}{W(x)}$ which after integration read
 \be
 a(x)=-\alpha^2\int dx\,x^2n_l(x),\qquad\qquad b(x)=\alpha^2\int dx\,x^2j_l(x).
 \ee
The particular solutions $\tilde f_p(x)$ are given in terms of spherical Bessel functions, 
the sine integral ${\rm Si}(x):=\int_0^xdt\frac{\sin t}{t}$ and the cosine integral ${\rm Ci}(x):=-\int_x^\infty dt\frac{\cos t}{t}$. First five particular solutions labeled by $l=0,\ldots,4$ have the form 
\begin{align}
\tilde f_p^{(l=0)}(x)&=\alpha^2,\\
\tilde f_p^{(l=1)}(x)&=\alpha^2\left(1+\frac{2}{x^2}\right),\\
\tilde f_p^{(l=2)}(x)&=\alpha^2\left(1+\frac{9}{x^2}\right)+3\alpha^2\left[\,{\rm Ci}(x)j_2(x)+\,{\rm Si}(x)n_2(x)\right],\\
\tilde f_p^{(l=3)}(x)&=\alpha^2\left(1+\frac{12}{x^2}+\frac{120}{x^4}\right),\\
\tilde f_p^{(l=4)}(x)&=\alpha^2\left(1+\frac{25}{2x^2}+\frac{1575}{2x^4}\right)+\frac{15\alpha^2}{2}\left[\,{\rm Ci}(x)j_4(x)+\,{\rm Si}(x)n_4(x)\right].
\end{align}

Solution with $l=0$ must take some non-zero value at the center $x=0$. This condition can be satisfied for ${\cal B}=0$. The remaining free parameters which are the constant ${\cal A}$ and the compacton radius $x_R$ can be determined from $f(x_R)=0$ and $f'(x_R)=0$. It gives $x_R= x_1^{1}$ where $x_1^{1}=4.49341$ is a first non-trivial zero of the spherical Bessel function $j_1(x)$. By trivial zero we mean $x_0^{1}=0$. The profile of the compacton is given by
\be
\tilde f(x)=\alpha^2\left(1-\frac{j_0(x)}{j_0(x_1^{1})}\right).\label{sol1}
\ee
Since $\alpha^2=\frac{\tilde \mu^2}{8\omega^2}$, then the radial profile function behaves as $\tilde f(x)=f(r)\propto \omega^{-2}$. From definition of the variable $x=\omega\, r$ one gets 
\be
R^{-1}=\frac{\omega}{x_1^{1}}\approx 0.22254\,\omega .
\ee
This formula allows to interpret the coefficient $A_1$ for $l=0$ in TABELE \ref{tab1a} as inverse of the first non-trivial  zero of $j_1(x)$.

For the model with $l=1$ the profile function reaches zero at $x=0$ and has non-vanishing first derivative at the center. It gives  ${\cal B}=2\alpha^2$ in \eqref{solgeral}. In order to satisfy boundary conditions at the compacton border one has to choose ${\cal A}=2\pi\alpha^2$ and $x_R=2\pi$. The solution is then of the form
\be
\tilde f(x)=\alpha^2\left(1+\frac{2}{x^2} +2\pi j_1(x)+2n_1(x)\right),\label{sol2}
\ee
where $j_1(x)=\frac{\sin(x)}{x^2}-\frac{\cos(x)}{x}$ and $n_1(x)=-\frac{\cos(x)}{x^2}-\frac{\sin(x)}{x}$.
The profile function is proportional to $\omega^{-2}$ and the compacton radius obeys relation
\be
R^{-1}=\frac{\omega}{2\pi}\approx 0.15915\,\omega,
\ee
which  allows to interpret $A_1$ for $l=1$ in TABELE \ref{tab1a} as $A_1=\frac{1}{2\pi} $. 

Let us consider the model with $l=2$. The  compacton radii $x_1$ and $x_2$ where $x_1<x_2$ are such that $\tilde f(x_1)=0$, $\tilde f'(x_1)=0$ and similarly  $ \tilde f(x_2)=0$, $\tilde f'(x_2)=0$. The boundary conditions at $x_1$ allows to determine constants ${\cal A}$ and ${\cal B}$ in \eqref{solgeral}. The solution takes the form
\begin{align}
\tilde f(x):=\alpha^2\left(1+\frac{9}{x^2}\right)&+\alpha^2\left(4\cos x_1 +x_1\sin x_1+3[{\rm Ci}(x)-{\rm Ci}(x_1)]\right)j_2(x)\nonumber\\
&+\alpha^2\left(4\sin x_1 -x_1\cos x_1+3[{\rm Si}(x)-{\rm Si}(x_1)]\right)n_2(x).\label{sol3}
\end{align}
The compacton radii $x_1$ and $x_2$ are determined by conditions $\tilde f(x_2)=0$, $\tilde f'(x_2)=0$. It gives $x_1=0.193871$ and $x_2=7.944507$ what leads to the compacton size  $\delta x=7.750640$. It follows that
\be
\delta R^{-1}=\frac{\omega}{\delta x}\approx 0.12902\,\omega.
\ee
 This result constitute quite good approximation of the coefficient $A_1$ for $l=2$, which is presented in TABELE \ref{tab1a}. Due to complexity of solution \eqref{sol3}, we cannot give an expression for coefficient $A_1$.

Finally, we shall discuss the relation between the energy and the Noether charges. The solutions \eqref{sol1},   \eqref{sol2}, and \eqref{sol3} have a form
\be
f(r)=\tilde f(x)=\alpha^2g(x),
\ee 
where $g(x)$ does not depend on $\omega$.

The energy density \eqref{sign-gordon-ed} can be cast in the form
\begin{align}
H(r)=\frac{\tilde \mu^4}{8\,\omega^2}\left[\frac{1}{2}\left( g'^2+\left(1+\frac{l(l+1)}{x^2}\right) g^2\right)+g\,{\rm sgn}\,(g)\right]\equiv\frac{\tilde \mu^4}{8\, \omega^2}G(x),
\end{align}
where $g'(x)=\frac{dg}{dx}(x)$. A total energy $E=4\pi\int_{0}^{\infty}dr\, r^2 H(r)$ reads
\begin{align}
E= {\varepsilon_1}\frac{\tilde\mu^4}{\omega^5},\qquad{\rm where}\qquad \varepsilon_1:=\frac{\pi}{2}\int_{0}^{\infty}dx\,x^2 G(x)\label{signGenergy}
\end{align}
is a numerical constant which does not depend on $\omega$. A contribution to $\varepsilon_1$ comes from the region where $G(x)$ is different from zero {\it i.e.} from the support $[0,x_2]$ for Q-balls and from $[x_1,x_2]$ for Q-shells. The proportionality  of $E^{-1/5}$ to $\omega$ in  \eqref{signGenergy} is a consequence of the relation $f(r)\propto \omega^{-2}$.

The Noether charges are given by expression
\begin{align}
 Q_t&=\varepsilon_2\frac{\tilde\mu^4}{\omega^6},\qquad {\rm where} \qquad \varepsilon_2:=\frac{\pi}{4(2l+1)}\int_0^{\infty}dx\,x^2 g^2\label{signGcharge}
\end{align}
is another numerical constant which does not depend on $\omega$.

In TABLE \ref{tab4} we present numerical values of coefficients $\varepsilon_1$ and $\varepsilon_2$. In particular, expressions $(\varepsilon_1)^{-1/5}$ are good approximations for coefficients  $A_2$ presented in TABLE \ref{tab1a}. Similarly, expressions $(\varepsilon_2)^{-1/6}$ are qualitatively good approximations of coefficients $A_3$ in TABLE \ref{tab2}. In case of Q-shells the concordance is not as good as for Q-balls.

\begin{table}[h!]
\centering
\label{my-label}
\begin{tabular}{|l||c|c||c|c|}
\hline
    & $\varepsilon_1$ & $(\varepsilon_1)^{-1/5}$ & $\varepsilon_2$ & $(\varepsilon_2)^{-1/6}$   \\ \hline\hline
$\;l=0\;$ & $\;142.511\;$  & $\;0.371\;$  & $\;59.379\;$  &  $\;0.506\;$    \\ \hline
$\;l=1\;$ & $508.072$  & $0.287$  & $70.565$  & $0.492$     \\ \hline
$\;l=2\;$ &  $1050.90$ &  $0.248$ & $160.813$  & $0.428$     \\ \hline
\end{tabular}\caption{Numerical constants $\varepsilon_1$ and $\varepsilon_2$.}\label{tab4}
\end{table}

The relations \eqref{signGenergy} and \eqref{signGcharge} imply that  relation between the energy and the Noether's charge is of the form
\be
E=\varepsilon_1\tilde\mu^{\frac{2}{3}}\left(\frac{Q_t}{\varepsilon_2}\right)^{\frac{5}{6}}.
\ee
The power $5/6$ suggest that the energy of two Q-balls (or Q-shells) with the charges $Q_1$ and $Q_2$ is higher than the energy of a single Q-ball (Q-shell) that has a charge $Q_1+Q_2$.

\section{Summary}

We have shown that the $CP^{2l+1}$ model with V-shaped potential possesses nontopological compact solutions with  finite energy in 3+1 dimensions. The solutions have the form of Q-balls for $l=0,1$ and Q-shells for $l\ge 2$. The Q-ball solution, $l=0$, is spherically symmetric, however, field configurations containing more than one scalar field are not. Note, that the energy density is spherically symmetric in all cases. The configuration of fields $u_m$ with $l\ge 1$ possesses some non-zero angular momentum. One can imagine that the existence of such angular momentum is associated with mutual motion of the fields $u_m$. It is consistent with the fact that the configuration containing a single scalar field has vanishing angular momentum $l=0$. The energy of the solutions is proportional to the Noether's charge in power  approximately $\frac{5}{6}$, what suggest that the solutions have no tendency to spontaneous decay into higher number of smallest Q-balls. This power is exact for solutions of the limit model obtained for $\omega\gg 1$. The limit model is recognized as the signum-Gordon type model which possesses a characteristic V-shaped non-linearity. Unlike for the original model, there is no lower bound for the parameter $\omega_c$ in the case of the signum-Gordon type model. In fact, all its solutions  are proportional to $\omega^{-2}$.  Although solutions of the signum-Gordon type model exist for all $\omega>0$, only those with $\omega\gg1$ are sufficiently close to solutions of the original  $CP^{2l+1}$ type model.

The compact solutions considered in this paper can be composed together so they form some multi Q-ball solutions. Such a composition is possible due to compactness of the individual solutions.  This property results in absence of interaction between individual Q-balls unless their supports overlap. Moreover, since the model possesses the Lorentz's symmetry, then acting with Lorentz boost on the solution describing Q-ball one gets a Q-ball in motion. Although we have not presented the explicit form of such solutions in this paper, it is quite straightforward that their construction can be performed in the same way as for compactons in the version of the model with two V-shaped minima \cite{chain}. 

This work can be continued in many directions, however, two of them seem to be essential. The first direction would be considering the $CP^N$ type models with even number of scalar fields. It requires an adequate ansatz which would allow to  reduce the $N$ equations of motion to a single radial equation. This problem is still open and requires some further studies. The second direction, which is our original motivation, is searching for compactons in the $CP^N$ SF type model with the potential. Our ansatz properly works for the model with odd number of scalar fields. An inclusion of further quartic terms in the Lagrangian would result in some new terms in the radial equation. With each such quartic term there is associated one coupling constant.  Consequently, the number of free parameters of the model would certainly increase.  This work is already in progress and we shall soon report on the results.

\section{Appendix}

{\small
\noindent Our ansatz leads to
\begin{align}
u^{\dagger}\cdot \partial_tu&=i\omega f^2& u^{\dagger}\cdot \partial_ru&=f'f & u^{\dagger}\cdot \partial_{\theta} u&=0& u^{\dagger}\cdot \partial_{\phi} u&=0\nonumber\\
\partial_{\theta}u^{\dagger}\cdot \partial_{\theta}u&=\frac{l(l+1)}{2}f^2& \partial_ru^{\dagger}\cdot \partial_ru&=f'^2& \partial_{t}u^{\dagger}\cdot \partial_{r}u&=-i\omega f'f&\partial_{\theta}u^{\dagger}\cdot \partial_{\phi}u &=0\nonumber\\
\partial_{\phi}u^{\dagger}\cdot \partial_{\phi}u&=\frac{l(l+1)}{2}\sin^2\theta f^2&\partial_{t}u^{\dagger}\cdot \partial_{t}u &=\omega^2 f^2&\partial_ru^{\dagger}\cdot \partial_{\alpha}u&=0&\partial_tu^{\dagger}\cdot \partial_{\alpha}u&=0\nonumber
\end{align}
}where $\alpha=\{\theta, \phi\}$. 

\section*{Acknowledgments}
The authors  are indebted to L.A. Ferreira and A. Wereszczy\'nski for discussion and valuable comments.

\end{document}